\documentstyle[pra,epsfig,aps]{revtex}
\input{psfig.sty}
\begin{document}

\title{Analysis of the computational complexity of solving random
satisfiability problems using branch and bound search algorithms.}

\author{Simona Cocco $^{1,2}$
and R{\'e}mi Monasson $^{1,3}$}

\address{$^1$ CNRS-Laboratoire de Physique Th{\'e}orique de l'ENS,
24 rue Lhomond, 75005 Paris, France, \\
$^{2}$ Department of Physics,
The University of Illinois at Chicago, 845 W. Taylor St., Chicago
IL 60607, USA,\\
$^{3}$ The James Franck Institute, The University of Chicago,
5640 S. Ellis Av., Chicago IL 60637, USA.}  
\date{\today}
\maketitle
\begin{abstract}
The computational complexity of solving 
random 3-Satisfiability (3-SAT) problems is investigated.
3-SAT is a representative example of
hard computational tasks; it consists in knowing whether a set of $\alpha
N$ randomly drawn logical constraints involving $N$ Boolean variables can be
satisfied altogether or not. Widely used solving procedures, as the 
Davis-Putnam-Loveland-Logeman (DPLL) algorithm, perform a systematic
search for a solution, through a sequence of trials and errors
represented by a search tree. The size of the search tree
accounts for the computational complexity, {\em i.e.} the amount of 
computational efforts, required to achieve resolution. 
In the present study, we identify, using theory and numerical experiments,
easy (size of the search tree scaling polynomially with $N$) 
and hard (exponential scaling) regimes  
as a function of the ratio $\alpha$ of constraints per variable.
The complexity is explicitly calculated in the different regimes,
in very good agreement with numerical simulations.
Our theoretical approach is based on the analysis of the growth of the
branches in the search tree under the operation of DPLL. 
On each branch, the initial 3-SAT problem is dynamically turned into
a more generic 2+p-SAT problem, where $p$ and $1-p$ are the fractions of 
constraints involving three and two variables respectively.
The growth of each branch is monitored by
the dynamical evolution of $\alpha$ and $p$ and is represented by a
trajectory in the static phase diagram of the random 2+p-SAT problem.
Depending on whether or not the trajectories cross the boundary between
satisfiable and unsatisfiable phases, single branches or full trees are 
generated by DPLL, resulting in easy or hard resolutions. 
Our picture for the origin of complexity can be applied to other
computational problems solved by branch and bound algorithms.
\end{abstract}
\vskip .5cm
PACS Numbers~: 05.10, 05.70, 89.80
\vskip .5cm                                                                     
\section{Introduction.} 

Out-of-equilibrium dynamical properties of physical systems form the 
subject of intense studies in modern statistical physics\cite{Gro}. 
Over the past decades, much progress has been made in fields as 
various as glassy dynamics, growth processes, persistence phenomena,
vortex depinning ... where dynamical aspects play a central role. 
Among all the questions related to these issues, the existence and
characterization of stationary states reached in some asymptotic limit
of large times is of central importance. In turn, the notion of asymptotic
regime raises the question of relaxation, or transient, behavior: what
time do we need to wait for in order to let the system relax? How
does this time grow with the size of the system? Such interrogations 
are not limited to out-of-equilibrium dynamics but also
arise in the study of critical slowing down phenomena accompanying
second order phase transitions.

Computer science is another scientific discipline where dynamical
issues are of central importance. There, the main question is to know
the time or, more precisely, the amount of computational resources
required to solve some given computational problem, and how this time
increases with the size of the problem to be solved\cite{NPC}. Consider for
instance the sorting problem\cite{Knu}. One is given a list ${\cal L}$ of $N$
integer numbers to be sorted in increasing order. What is the
computational complexity of this task, that is, the minimal number of
operations (essentially comparisons) necessary to sort any list ${\cal
L}$ of length $N$? Knuth answered this question in the early
seventies: complexity scales at least as $N \log N$ and there exists a
sorting algorithm, called Mergesort, achieving this lower
bound\cite{Knu}. 

To calculate computational complexity, one has to study how the
configuration of data representing the computational problem
dynamically evolves under the prescriptions encoded in the
algorithm. Let us consider the sorting problem
again and think of the initial list ${\cal L}$ as a (random)
permutation of ${\cal I} = \{1,2, \ldots , N\}$. Starting from ${\cal
L} (0) = {\cal L}$, at each time step (operation) $T$, the sorting
algorithm transforms the list ${\cal L} (T)$ into another list ${\cal
L} (T+1)$, reaching finally the identity permutation, {\em i.e.} the
ordered list ${\cal I}$. Obviously, the dynamical rules imposed by a
solving algorithm are of somewhat unusual nature from a physicist's
point of view. They might be highly non-local and non-Markovian. Yet,
the operation of algorithms gives rise to well posed dynamical
problems, to which methods and techniques of statistical physics may
be applied as we argue in this paper.

Unfortunately, not all problems encountered in computer science are as
simple as sorting. Many computational problems originating from
industrial applications, e.g. scheduling, planning and more generally
optimization tasks, demand computing efforts growing enormously with
their size $N$. For such problems, called NP--complete, all known
solving algorithms have execution times increasing {\em a priori}
exponentially with $N$ and it is a fundamental conjecture of computer
science that no polynomial solving procedure exists. To be more
concrete, let us focus on the 3-satisfiability (3-SAT) problem, a
paradigm of the class of NP--complete computational
problems\cite{NPC}. A pedagogical introduction to the 3-SAT problem
and some of the current open issues in theoretical computer science
may be found in~\cite{Hay}.  

3-SAT is defined as follows.  Consider a set of $N$ Boolean variables
and a set of $M=\alpha\, N$
constraints (called clauses), each of which being the logical OR of
three variables or of their negation. Then, try to figure out whether
there exists or not an assignment of variables satisfying all clauses.
If such a solution exists, the set of clauses (called instance of the
3-SAT problem) is said satisfiable (sat); otherwise the instance is
unsatisfiable (unsat).  To solve a 3-SAT instance, {\em i.e.} to know
whether it is sat or unsat, one usually resorts to search algorithms,
as the ubiquitous Davis--Putnam--Loveland--Logemann (DPLL) 
procedure\cite{Hay,DP,Cra}. DPLL operates by trials and errors, the
sequence of which can be graphically represented as a search
tree. Computational complexity is the amount of operations performed
by the solving algorithm and is conventionally measured by the size of
the search tree.

Complexity may, in practice, vary enormously with the instance of the
3-SAT problem under consideration.  To understand why instances are
easy or hard to solve, computer scientists have focused on model
classes of 3-SAT instances.  Probabilistic models, that define
distributions of random instances controlled by few parameters, are
particularly useful.  An example, that has attracted a lot of
attention over the past years, is random 3-SAT: all clauses are drawn
randomly and each variable negated or left unchanged with equal
probabilities.  Experiments\cite{Cra,Che,Mit,AI,Kir} and
theory\cite{Gut,MZ} indicate that instances are almost surely always
sat (respectively unsat) if $\alpha$ is smaller (resp. larger) than a
critical threshold $\alpha _C \simeq 4.3$ as soon as $M,N$ go to
infinity at fixed ratio $\alpha$. This phase transition\cite{MZ,Sta}
is accompanied by a drastic peak of computational hardness at
threshold\cite{Cra,Che,Mit}, see Figure~\ref{tempi}. Random
3-SAT generates simplified and idealized versions of real-world
instances. Yet, it reproduces essential features (sat vs. unsat, easy
vs. hard) and can shed light on the onset of complexity, in the same
way as models of condensed matter physics help to understand global
properties of real materials.

Phases in random 3-SAT, or in physical systems, characterize the
overall {\em static} behavior of a sample in the large size limit --
a large instance with ratio e.g. $\alpha =3$ will be almost surely sat
(existence proof) -- but do not convey direct information of {\em
dynamical} aspects -- how long it will take to actually find a
solution (constructive proof).  This situation is reminiscent of the
learning problem in neural networks (``equilibrium'' statistical
mechanics allows to compute the maximal storage capacity, irrespective
of the memorization procedure and of the learning time)
\cite{Neur}, or liquids at
low enough temperatures (that should crystallize from a
thermodynamical point of view but undergo some kinetical glassy
arrest)\cite{Ange}\footnote{The analogy between relaxation 
in physical systems and
computational complexity in combinatorial problems is even clearer when
the latter are solved using local search algorithms, e.g. simulated
annealing or other solving strategies making local moves based on some
energy function.  Consider a random 3-SAT instance below threshold. We
define the energy function (depending on the configuration of Boolean
variables) as the number of unsatisfied clauses\cite{Walk}. The goal of the
algorithm is to find a solution, {\em i.e.}  to minimize the
energy. The configuration of Boolean variables evolve from some
initial value to some solution under the action of the
algorithm. During this evolution, the energy of the ``system'' relaxes
from an initial (large) value to zero. Computational complexity is,
in this case, equal to the relaxation time of the dynamics before
reaching (zero temperature) equilibrium.}.

This paper is an extended version of a previous work \cite{Let}, where
we showed how the dynamics induced by the DPLL search
algorithm could be analyzed using off-equilibrium statistical mechanics
and combined to the static phase diagram of random K-SAT (with
K=2,3) to calculate
computational complexity. We start by exposing in a detailed way the
definition of the random K-SAT problem and the DPLL procedure in
Section~\ref{secdpll0}. We then expose the experimental measures of
complexity in Section~\ref{secnum}. Our analytical approach is based on
the fact that, under DPLL action, the initial instance is modified and
follows some trajectory in the phase diagram.  The structure of the
search tree generated by DPLL procedure is closely related to the
nature of the region visited by the instance trajectory.  Search trees
reduce to essentially one branch -- sat instances at low ratio
$\alpha$, section~\ref{secbranch} -- or are dense, including 
an exponential number of
branches -- unsat instances, section~\ref{sectree}. Mixed structures
-- sat instances with ratios slightly below threshold,
section~\ref{secmixed} -- are made of a branch followed by a dense
tree and reflect trajectories crossing the phase boundary between sat
and unsat regimes. While branch trajectories could be obtained
straightforwardly from previous works by Chao and Franco\cite{Fra}, 
we develop in section~\ref{sectree}
a formalism to study the construction of
dense trees by DPLL. We show that the latter can be reformulated in
terms of a (bidimensional) growth process described by a non-linear
partial differential equation.  
The resolution of this growth equation allows an
analytical prediction of the complexity that compares very well to
extensive numerical experiments. We  present in Section~\ref{secpattern}
the full complexity diagram of solving random SAT models, and explain 
the relationship with static studies of the phase transition \cite{Sta}.  
Last of all, we show in Section~\ref{secconc} how our study suggests 
some possible ways to improve existing algorithms.

\section{Davis-Putnam-Loveland-Logeman algorithm and random 3-SAT.} 
\label{secdpll0}

In this section, the reader will be briefly recalled the main features
of the random 3-Satisfiability model. We then present the
Davis-Putnam-Loveland-Logeman (DPLL) solving procedure, a paradigm of branch
and bound algorithm, and the notion of search tree. Finally, we introduce the
idea of dynamical trajectory, followed by an instance under the action
of DPLL. 

\subsection{A reminder on random Satisfiability.} \label{reminder}

Random K-SAT is defined as follows.  Let us consider $N$ Boolean
variables $x_i$ that can be either true (T) or false (F)
($i=1,\ldots,N)$.  We choose randomly $K$ among the $N$ possible
indices $i$ and then, for each of them, a literal, that is, the
corresponding $x_i$ or its negation $\bar x_i$ with equal
probabilities one half. A clause $C$ is the logical OR of the $K$
previously chosen literals, that is $C$ will be true (or satisfied) if
and only if at least one literal is true.  Next, we repeat this
process to obtain $M$ independently chosen clauses
$\{C_\ell\}_{\ell=1,\ldots,M}$ and ask for all of them to be true at
the same time ({\em i.e.} we take the logical AND of the $M$ clauses).  The
resulting logical formula is called an instance of the K-SAT problem.
A logical assignment of the $x_i$'s satisfying all clauses, if any, is
called a solution of the instance.

For large instances ($M,N \to \infty$), K-SAT exhibits a striking
threshold phenomenon as a function of the ratio $\alpha =M/N$ of the
number of clauses per variable.  Numerical simulations indicate that
the probability of finding a solution falls abruptly from one down to
zero when $\alpha$ crosses a critical value 
$\alpha_C(K)$\cite{Cra,Che,Mit}.  Above
$\alpha_C(K)$, all clauses cannot be satisfied any longer.  This
scenario is rigorously established in the $K=2$ case, where $\alpha
_C=1$\cite{exact}.  For $K\ge3$, much less is known; $K(\ge 3)$--SAT
belongs to the class of hard, NP-complete computational
problems\cite{NPC}.  Studies have mainly concentrated on the $K=3$ case,
whose instances are simpler to generate than for larger values of
$K$. Some lower \cite{Fri} and upper \cite{Upper} 
bounds on $\alpha _C(3)$ have been
derived, and numerical simulations have recently allowed to
find precise estimates of $\alpha _C$, e.g. $\alpha _C (3) \simeq
4.3$\cite{Mit,Kir}.

The phase transition taking place in random 3-SAT has attracted a
large deal of interest over the past years due to its close 
relationship with the emergence of computational complexity. Roughly
speaking, instances are much harder to solve  at
threshold than far from criticality \cite{Cra,Che,Mit,AI}. 
We now expose the solving procedure used to tackle the 3-SAT problem.

\subsection{The Davis-Putnam-Loveland-Logeman solving procedure.} 
\label{secdpll}

\subsubsection{Main operations of the solving procedure and search trees.}

3-SAT is among the most difficult problems to solve as its size $N$
becomes large.   In practice, one
resorts to methods that need, {\em a priori}, exponentially
large computational resources. One of these algorithms, the 
Davis--Putnam--Loveland--Logemann (DPLL) solving
procedure\cite{Hay,DP}, is illustrated on Figure~1. DPLL operates
by trials and errors, the sequence of which can be graphically
represented as a search tree made of nodes connected through edges as
follows:
\begin{itemize}
\item[1.] A node corresponds to the choice of a variable. Depending
on the value of the latter, DPLL takes one of the two possible
edges. 
\item[2.] Along an edge, all logical implications of the last choice
made are extracted. 
\item[3.] DPLL goes back to step~1 unless a solution is
found or a contradiction arises; in the latter case, DPLL backtracks
to the closest incomplete node (with a single descendent edge),
inverts the attached variable and goes to step~2; if all nodes carry
two descendent edges, unsatisfiability is proven.
\end{itemize}

Examples of search trees for satisfiable (sat) or unsatisfiable
(unsat) instances are shown Figure~\ref{tree}. Computational complexity is the
amount of operations performed by DPLL, and is measured by the size of 
the search tree, {\em i.e.} the number of nodes.

\subsubsection{Heuristics of choice.} \label{heuri}

In the above procedure, step~1 requires to choose one literal among
the variables not assigned yet. The choice of the variable and of its
value obeys some more or less empirical rules called
splitting heuristics. The key idea is to choose variables that will
lead to the maximum number of logical implications \cite{Cook}.
Here are some simple heuristics:
\begin{itemize}
\item {\em ``Truth table'' rule:} fix unknown variables in lexicographic
order, from $x_1$ up to $x_N$ and assign them to e.g. true. This is an
inefficient rule that does not follow the key principle exposed above. 

\item {\em Generalized Unit-Clause (GUC) rule:} choose randomly one
literal among the shortest clauses\cite{Fra}. This is an extension of
unit-propagation that fixes literal in unitary clauses. GUC is based
on the fact that a clause of length $K$ needs at most $K-1$ splittings to 
produce a logical implication. So variables are chosen preferentially 
among short clauses.

\item {\em Maximal occurrence in minimum size clauses (MOMS) rule:}
pick up the literal appearing most often in shortest clauses. This
rule is a refinement of GUC. 
\end{itemize}

Global performances of DPLL depend quantitatively on the splitting
rule.  From a qualitative point of view, however, the easy-hard-easy
picture emerging from experiments is very robust
\cite{Che,Mit,Kir}.  Hardest instances seem to be located
at threshold. Solving them demand an exponentially large computational
effort scaling as $2^{N\omega_C}$. The values of $\omega _C$ found in
literature roughly range from $0.05$ to $0.1$, depending on the
splitting rule used by DPLL\cite{Cra,Cook}.

In this paper, we shall focus on the GUC heuristic which is simple
enough to allow analytical studies and, yet, is already quite efficient.

\subsection{2+p-SAT and instance trajectory.} \label{deuxpp}

We shall present in Section~\ref{secnum} the experimental results on solving
3-SAT instances using DPLL procedure in a detailed way. The main
scope of this paper is to compute in an analytical way the 
computational complexity $\omega$ in the easy and hard regimes. 
To do so, we have made use of the precious notion of
dynamical trajectory, that we now expose.

As shown in Figure~\ref{algo}, the action of DPLL on an instance of
3-SAT causes the reduction of 3-clauses to 2-clauses. Thus, a mixed
2+p-SAT distribution\cite{Sta}, where $p$ is the fraction of
3-clauses, may be used to model what remains of the input instance at
a node of the search tree. A 2+p-SAT formula of parameters $p,\alpha$ 
is the logical AND of two uncorrelated random 2-SAT and 3-SAT instances
including $\alpha\, (1-p)\, N$ and $\alpha\, p\, N$ clauses respectively.
Using experiments\cite{Sta} and statistical mechanics calculations\cite{MZ}, 
the threshold line $\alpha _C (p)$ may be obtained with the results 
shown in Figure~\ref{diag} (full line). Replica calculations 
suggest that the sat/unsat transition taking place at $\alpha _C (p)$ is
continuous if $p <p_S$ and discontinuous if $p>p_S$, where $p_S \simeq 
0.41$\cite{Tric}.
The tricritical point T$_S$ is shown in Figure~\ref{diag}. Below
$p_s$, the threshold $\alpha _C (p)$ coincides with the upper bound $1/(1-p)$,
obtained when requiring that the 2-SAT subformula only be satisfied.
Rigorous studies have shown that $p_S\ge 2/5$, leaving
open the possibility it is actually equal to 2/5\cite{Varia}, 
or to some slightly larger value\cite{Achl}. 

The phase diagram of 2+p-SAT is the natural space in which DPLL
dynamic takes place\cite{Let}. 
An input 3-SAT instance with ratio $\alpha _0$ shows
up on the right vertical boundary of Figure~\ref{diag} as a point of
coordinates $(p=1,\alpha _0 )$.  Under the action of DPLL, the
representative point moves aside from the 3-SAT axis and follows a
trajectory. The location of this trajectory in the phase diagram
allows a precise understanding of the search tree structure and of
complexity.

\section{Numerical Experiments.} \label{secnum}

\subsection{Description of the numerical implementation of the DPLL algorithm.}

We have implemented DPLL with the GUC rule, see Figure~\ref{algo} and 
Section~\ref{heuri}, to have a fast unit
propagation and an inexpensive backtracking \cite{Cra}. The program is
divided in three parts. The first routine draws the
clauses and  represents the data in a convenient structure.  The second,
main routine updates and saves the state of the search, {\em i.e.}  the
indices and values of assigned variables, to allow an
easy backtracking. Then, it checks if a solution is found; if not,
a new variable is assigned. 
The third routine extracts the implication of the choice (propagation).
If unit clauses have been generated, the corresponding literals are
fixed, or a contradiction is detected.
 
\subsubsection{Data Representation}

Three arrays are used to encode the data: the two first arrays are labelled
by the clause number $m=1,\ldots, M$ and the number $b=1,\ldots, K$ of 
the components in the clause (with $K=2,3$). 
The entries of these arrays, initially drawn at random, are the indices 
$clausnum\, (m,b)$ and  the values $ clausval\,(m,b)$,
true or false, of the variables. The indices of the third array are 
integers $i=1,\ldots ,N$ and $j=1,\ldots , o_i$ where $o_i$ is the number 
of occurrences of $x_i$  in the clauses (from zero to $M$). 
The entries of the matrix, $a\,(i,j)$,  are the numbers of the corresponding
clauses (between 1 and $M$).

\subsubsection{Updating of the search state.}

If the third routine has found a contradiction,  the second routine
goes back along the branch, inverts the last assigned variable
and calls again the third routine. If not, the descent along the branch 
is pursued. A Boolean-valued vector points to
the assigned variables, while the values are stored
in another unidimensional array. 
For each clause, we check if the variables are already assigned and, if so, 
if they are in agreement or not with the clauses. When splitting occurs,
a new variable is fixed to satisfy a 2-clause, {\em i.e.} a clause with 
one false  literal and two unknown variables, and the third subroutine  
is called. If there are only 3-clauses, a
new variable is fixed to satisfy any 3-clause and
the third subroutine is called. The variable
chosen and its value are stored in a vector with index the length of the
branch, {\em i.e.} the number of nodes it contains, to allow
future backtracking. If there are neither 2- nor 3-clauses left, a solution is
found. 
 
\subsubsection{Consequences of a choice and unit propagation}

All clauses containing the last fixed variable  
are analyzed by taking into account all possibilities: 1. the clause
is satisfied; 2. the clause is reduced to a 2- or 1-clause; 3. the
clause is violated (contradiction). In the second case, the 1-clause is stored
to be analyzed by unit-propagation once all clauses containing 
the variable have been reviewed.

\subsection{Characteristic running times}

We have implemented the DPLL search algorithm in Fortran~77;
the algorithm runs on a Pentium II PC with a 433M Hz frequency clock.
The number of nodes added per minute ranges from 
300,000 (typically obtained for $\alpha=3.5$) to
100,000 ($\alpha=10$) since unit propagation is more and more 
frequent as $\alpha$ increases.
The order of magnitude of the computational time needed to solve an instance
are listed in Table~\ref{tabtemps} for ratios corresponding to hard instances. 
These times limit the maximal size $N$ of the instances we have
experimentally studied and the number of samples over which we have 
averaged. Some rare instances
may be much harder than the typical times indicated in Table~\ref{tabtemps}. 
For instance, for $\alpha=3.1$ and $N=500$, instances
are usually solved in about 4 minutes but some samples required
more than 2 hours of computation. Such a phenomenon will be discussed
in Section~\ref{foc}.

\subsection{Overview of experiments }

\subsubsection{Number of nodes of the search tree}

We have first investigated complexity by a direct  count of
the number of  splittings, that is the number of nodes (black 
points) in Figure~\ref{tree}, for sat (Figure~\ref{tree}A,C) and
unsat (Figure~\ref{tree}B) trees.

\subsubsection{Histogram of branch length.}

We have also experimentally investigated  the structure
of search trees for unsat instances (Figure~\ref{tree}B). 
A  branch is defined as a path joining the root (first
node on the top of the search tree) to a leaf  marked 
with a contradiction $C$ (or a solution $S$ for sat
instance) in Figure~\ref{tree}.  
The length of a branch is the number of nodes it contains.
For an unsat instance, the complete search tree depends on 
the variables chosen at each split, but not on the values they are assigned to.
Indeed, to prove that there is no solution, all nodes in the search tree
have to carry two outgoing branches,  
corresponding to the two choices of the attached variables. What
choice is made first does not matter. 
This simple remark will be of  crucial importance in the theoretical analysis
of Section~\ref{paral}. 
 
We have derived the histogram of the branch lengths by counting
the number $B(l)$ of branches having length $l \,N$ once the tree
is entirely built up. The histogram is very useful to deduce the 
complexity in the unsat phase, since in a complete tree the
total number of branches $B$ is related  to the number of nodes $Q$
through the identity, 
\begin{equation}
\label{eqnb}
B=\sum _{l=\frac 1N , \ldots , 1} B(l) = Q+1 \qquad .
\end{equation}
that can be inferred from Figure~\ref{tree}B.

\subsubsection{Highest backtracking point} \label{high}

Another key property we have focused upon is the
highest backtracking point in the search tree.
In the unsat phase, DPLL backtracks all the nodes of the tree since no
solution can be present. The highest backtracking point in the tree
simply coincides with the top (root) node. In the sat phase, the
situation is more involved. A solution generally requires some
backtracking and the highest backtracking node may be defined as the
closest node to the origin through which two branches pass, node $G$ on
Figure~\ref{tree}B. We experimentally keep trace of the
highest backtracking point by measuring the numbers $C_2 (G)$, $C_3(G)$
of 2- and 3-clauses, the number of not-yet-assigned variables
$N (G)$, and  computing the coordinates 
$p_G=C_3(G)/(C_2(G)+C_3(G)), \alpha_G=(C_2 (G)+C_3 (G))/N(G)$ 
of $G$ in the phase diagram of Figure~\ref{diag}.

\subsection{Experimental Results}

\subsubsection{Fluctuations of complexity.}
\label{foc}
The size of the search tree built by DPLL is a random variable, due to
the (quenched) randomness of the 3-SAT instance and the choices made
by the splitting rule (``thermal noise''). We show on
Figure~\ref{histoseuil} the distribution of the logarithms (in base 2, and
divided by $N$) of the number of nodes for different values
of $N$. The distributions are more and more peaked around their mean
values $\omega _N (\alpha )$ as the size $N$ increases. This indicates
that the logarithm of the complexity is a self-averaging quantity in
the thermodynamic limit.  However, fluctuations are dramatically
different at low and large ratios.  For $\alpha =10$, and more
generally in the unsat phase, the distributions are roughly
symmetric (Figure~\ref{histoseuil}A).  Tails are small and
complexity does not fluctuate too much from sample to
sample\cite{Kirk2,Easy}. In the vicinity of $\alpha_L$, e.g. $\alpha
=3.1$, much bigger fluctuations are present. There are large tails on
the right flanks of the distributions on Figure~\ref{histoseuil}B, due
to the presence of rare and very hard samples \cite{Easy}. Complexity is not
self-averaging. We will come back to this point in
section~\ref{secmixed}.

\subsubsection{The easy-hard-easy pattern.}

\begin{center}
\begin{figure}
\includegraphics[width=235pt,height=350pt,angle=-90]{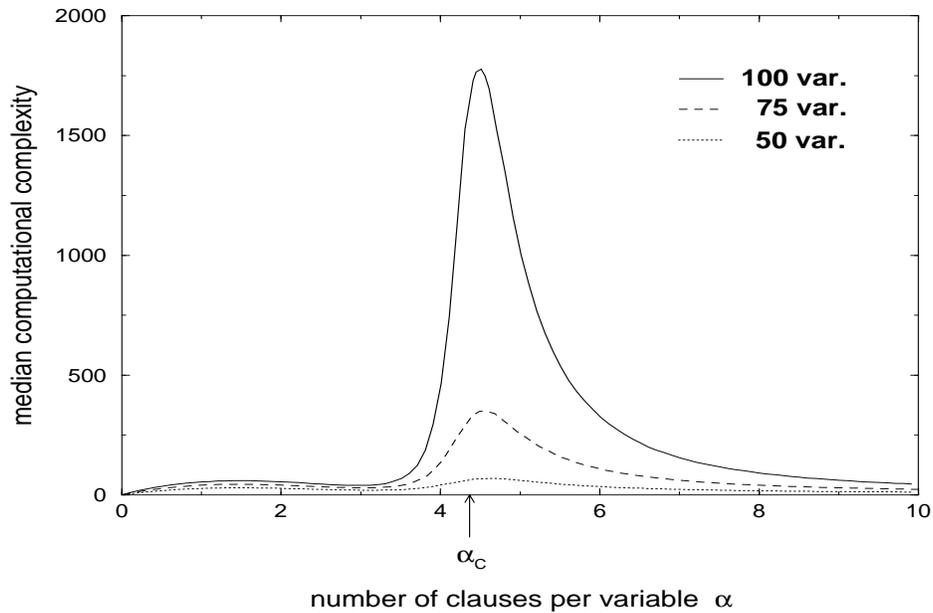}
\vskip 1cm
\caption{Solving complexity of random 3-SAT as a function of the ratio 
$\alpha$ of clauses per variables, and for three
increasing problem sizes $N$. Data are averaged over 10,000 randomly drawn
samples. Complexity is maximal at the threshold $\alpha _C\simeq 4.3$.}
\label{tempi}
\end{figure}
\end{center}

\begin{center}
\begin{figure}
\includegraphics[width=235pt,height=350pt,angle=-90]{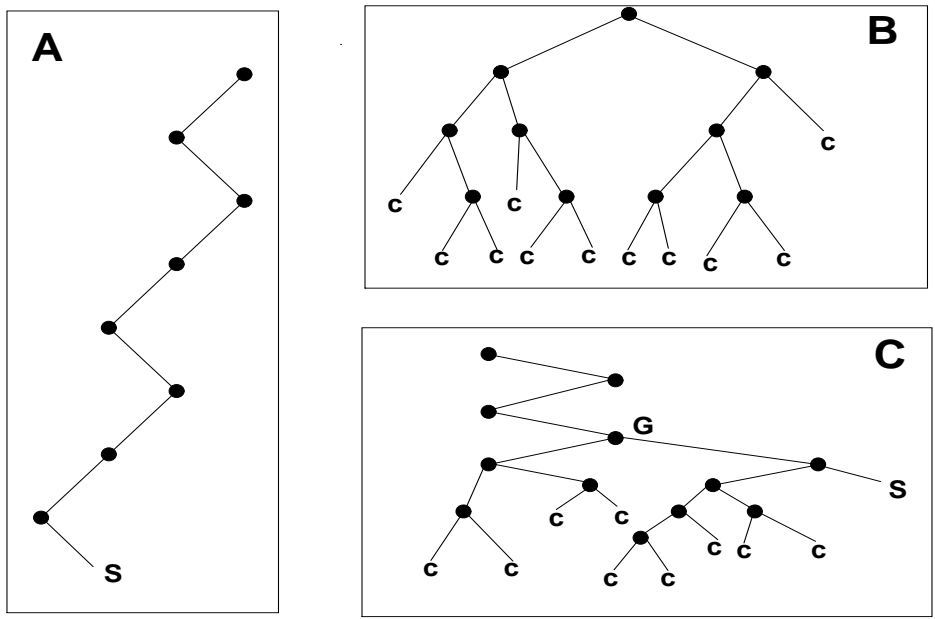}
\vskip 1cm
\caption{Examples of search trees.
{\bf A.} {\em simple branch:} the algorithm finds
easily a solution without ever (or with a negligible amount of) 
backtracking. {\bf B.} {\em dense tree:}
in the absence of solution, the algorithm builds a ``bushy'' tree,
with many branches of various lengths, before stopping.  {\bf C.} {\em
mixed case, branch + tree:} if many contradictions arise before
reaching a solution, the resulting search tree can be decomposed in a
single branch followed by a dense tree. The junction G is the highest
backtracking node reached back by DPLL.}
\label{tree}
\end{figure}
\end{center}

\begin{center}
\begin{figure}
\includegraphics[height=450pt,angle=0]{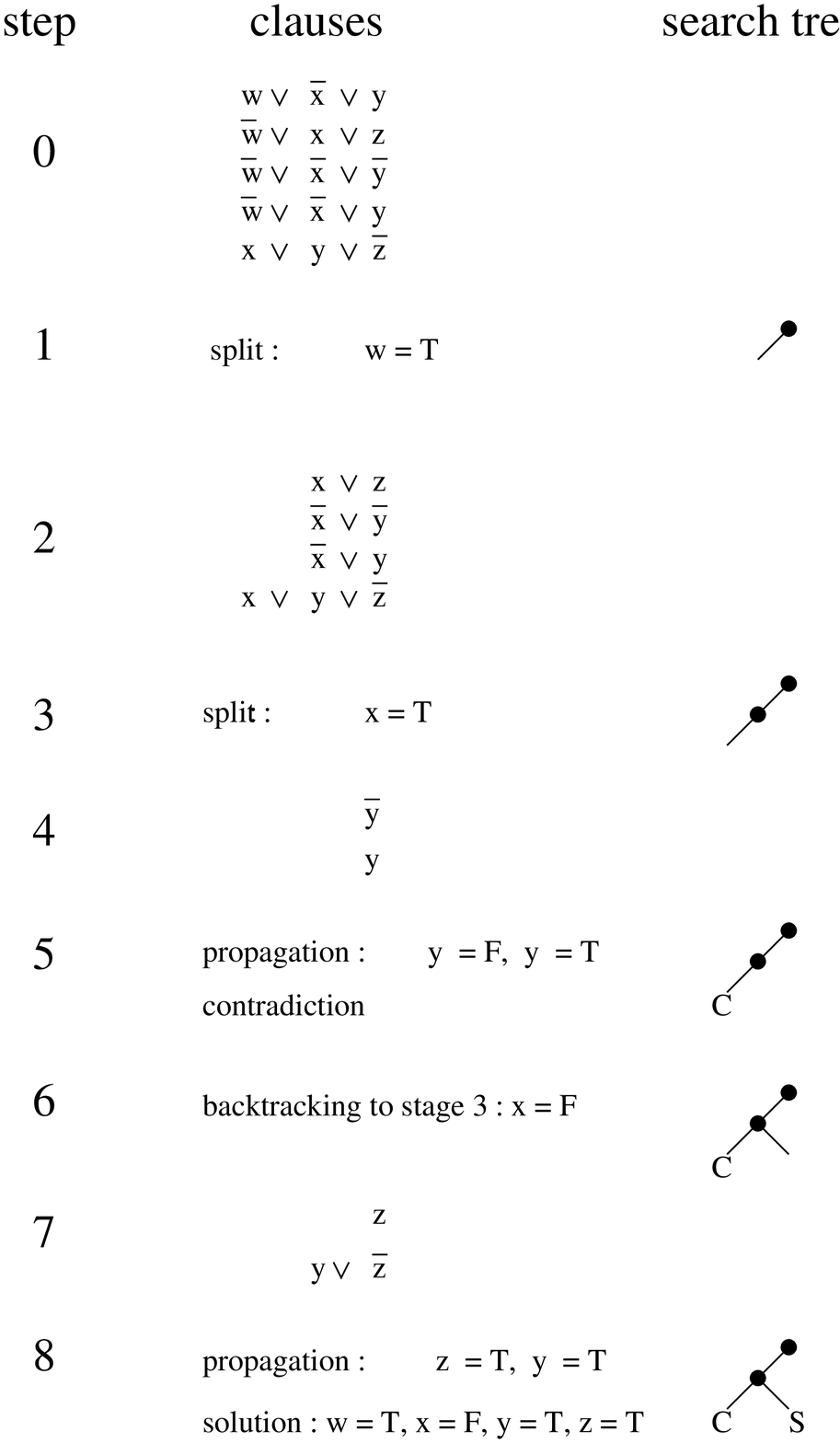}
\vskip 1cm
\caption{ Example of 3-SAT instance and Davis-Putnam-Loveland-Logemann
resolution. {\bf Step~0.}  The instance consists of $M=5$ clauses
involving $N=4$ variables that can be assigned to true (T) or false
(F). $\bar w$ means (NOT $w$) and v denotes the logical OR. The search
tree is empty.  {\bf 1.}  DPLL randomly selects a variable among the
shortest clauses and assigns it to satisfy the clause it belongs to,
e.g. $w=$T (splitting with the Generalized Unit Clause --GUC--
heuristic). A node and an edge symbolizing respectively
the variable chosen ($w$) and its value (T) are added to the tree.
{\bf 2.}  The logical implications of the last choice are extracted:
clauses containing $w$ are satisfied and eliminated, clauses including
$\bar w$ are simplified, and the remaining clauses are left unchanged. If
no unitary clause ({\em i.e.} with a single variable) is present, a
new choice of variable has to be made.  {\bf 3.}  Splitting takes
over. Another node and another edge are added to the tree.  {\bf 4.}
Same as step 2 but now unitary clauses are present.  The variables
they contain have to be fixed accordingly (propagation).  
{\bf 5.}  The propagation
of the unitary clauses results in a contradiction. The current branch
dies out and gets marked with C.  {\bf 6.}  DPLL backtracks to the
last split variable ($x$), inverts it (F) and creates a new edge.
{\bf 7.}  Same as step 4.  {\bf 8.}  Propagation of the unitary
clauses eliminates all the clauses. A solution S is found.
This example show how DPLL find a solution for a satisfiable instance.
For an unsatisfiable instance, unsatisfiability is proven when
backtracking (see step 6) is not possible anymore since all split
variables have already been inverted. In this case, all the nodes in
the final search tree have two descendent edges and all branches
terminate by a contradiction C.}
\label{algo}
\end{figure}
\end{center}

We have averaged the logarithm of the number of nodes over 
10,000 randomly drawn instances to obtain $\omega _N (\alpha)$. 
The  typical size $Q$ of the search tree is simply $Q=2^{N \omega _N}$. 
Results are shown in Figure~\ref{tempi}. An easy-hard-easy pattern of 
complexity appears clearly as the ratio $\alpha$ varies.

\begin{itemize}
\item At small ratios, complexity increases as $\gamma (\alpha ) N$,
that is, only linearly with the size of the instance.
DPLL easily finds a solution and the search tree
essentially reduces to a single branch shown on Figure~\ref{tree}A. 
For the GUC heuristic, the linear regime extends up to $\alpha _L
\simeq 3.003$ \cite{Fra,Fri}.

\item Above threshold, complexity grows exponentially
with $N$\cite{Chv}. The logarithm $\omega
(\alpha )$, limit of $\omega _N (\alpha)$ as $N \to \infty$, is maximal at
criticality...

\item ... and decreases at large ratios as $1/\alpha$ \cite{Bea}. 
The ``easy'' region on the right
hand side of Figure~\ref{tempi} is still exponential but with 
small coefficients $\omega$.
\end{itemize}

Of particular interest is the intermediate region $\alpha _L < \alpha
< \alpha _C$. We shall show that complexity is exponential in this
range of ratios, and that the search tree is a mixture of the
search trees taking place in the other ranges $\alpha < \alpha _L$
and $\alpha > \alpha _C$.

Let us mention that, while this paper is devoted to typical-case
(happening with probability one) complexity, rigorous results have 
been obtained that apply to any instance. So far, using a refined
version of DPLL, any instance is guaranteed to be solved in less than
$1.504^N$ steps, {\em i.e.} $\omega < 0.588$. The reader is referred
to reference \cite{Kul} for this worst-case analysis.
  
\subsubsection{Lower sat phase ($\alpha<\alpha_L$)}
\label{secnumlin}

The complexity data of Figure~\ref{tempi} obtained for different
sizes $N=50, 75, 100$ are plotted again on Figure~\ref{constante} after 
division by $N$. Data collapse on a single curve, proving that
complexity is linear in the regime $\alpha < \alpha _L$. 
In the vicinity of the cross over ratio $\alpha =\alpha_L$
 finite-size effects became important in this
region. We have thus performed extra simulations for larger sizes 
$N=500, 1000$ in the range $2.5 \le \alpha \le 3.$ that confirm the
linear scaling of the complexity.

\subsubsection{Unsat Phase ($\alpha>\alpha_C$)}

Results for the shape of the search trees are shown in
Figure~\ref{histo}.  We represent the logarithm $b(l)$, in base 2 and divided
by $N$, of the number $B(l)$ of branches as a function of the branch length
$l$, averaged over many samples and for different sizes $N$ and ratios 
$\alpha.$ When $\alpha$ increases 
at fixed $N$, branches are shorter and shorter and less and less
numerous, making complexity decrease (Figure~\ref{tempi}).

As $N$ gets large at fixed $\alpha$, 
the histogram $b(l)$ becomes a smooth function
of $l$ and we can replace the discrete sum in (\ref{eqnb}) with
a continuous  integral on the length,
\begin{equation}
Q+1=\sum_{l=\frac 1N,\ldots,1}  2^{N\,b(l)}
\simeq N\int_{0}^{1} dl \;
2^{N\,b(l)} \qquad . \label{int741}
\end{equation}
The integral is exponentially dominated by  
the maximal value $b_{max}$ of $b(l)$. $\omega$, the limit of the logarithm
of the complexity divided by $N$, is therefore equal to $b_{max}$.
Nicely indeed, the height $b_{max}$ of the histogram does not depend on $N$
(within the statistical errors)
and gives a straightforward and precise estimate of $\omega$, not
 affected by finite-size effects. The values of $b_{max}$ as a function
of $\alpha$ are listed in the third column of Table~\ref{tabunsat}.

The above discussion is also very useful to interpret the data
on the size $Q$ of the search trees. 
From the quadratic correction around the saddle-point,
$b(l) \simeq b_{max} -  \beta (l-l_{max})^2 /2 $, 
the expected finite size correction to $\omega _N$ read
\begin{equation}
\label{nodeq}
\omega _N  \simeq \omega + \frac 1{2N} \log_2 N +
\frac{1}{2N} \log_2 \left[ \frac{2\pi}{\beta \ln 2} \right] + O \left(
\frac 1 {N^2} \right) \qquad .
\end{equation}  
We have checked the validity of this equation by fitting
$\omega_N - \log_2 N / (2N)$ as a polynomial function of $1/N$.
The constant at the origin gives values of $\omega$ in very good
agreement with $b_{max}$ (second column in Table~\ref{tabunsat})
while the linear term gives access to the curvature $\beta$. 
We compare in Table~\ref{tabfit} this curvature with the direct
measurements of $\beta$ obtained by looking at the vicinity of the
top of the histogram. The agreement is fair, proving that equation 
(\ref{nodeq}) is an accurate way of extrapolating data on $Q$ to the
infinite size limit.

\subsubsection{Upper sat phase ($\alpha _L <\alpha < \alpha _C$)}

To investigate the sat region slightly below threshold $\alpha _L <
\alpha  < \alpha _C$, we have carried out simulations with a starting
ratio $\alpha  =3.5$. Results are shown on Figure~\ref{comp3.5}A.
As instances are sat with a high probability, no simple
identity relates the number of nodes $Q$ to the number of branches $B$,
see search tree in Figure~\ref{tree}C and
we measure the complexity through $Q$ only. Complexity clearly scales 
exponentially with the size of the instance
and exhibits large fluctuations from sample to sample. The annealed
complexity (logarithm of the average complexity), $\omega _3 ^{ann}$, 
is larger than the typical solving hardness (average of the logarithm 
of the complexity), $\omega _3 ^{typ}$, see Table~\ref{tabmixed}.

To reach a better understanding of the structure of the search tree,
we have focused on the highest backtracking point G defined in
Figure~\ref{tree}C and Section~\ref{high}. The coordinates $p_G, \alpha _G$
of point G, averaged over instances are shown for increasing sizes
$N$ on Figure~\ref{pgalphag}. The coordinates of G exhibit strong
sample fluctuations which make the large $N$ extrapolation, $p_G
=0.78 \pm 0.01$, $\alpha _G = 3.02 \pm 0.02$ rather imprecise.

In Section~\ref{secmixed}, we shall show how the solving complexity in the
upper sat phase is related to the solving complexity of corresponding
2+p-SAT problems with parameters $p_G, \alpha _G$.

\section{Branch Trajectories and the linear 
regime (lower sat phase).} \label{secbranch}

In this section, we investigate the dynamics of DPLL in the low ratio
regime, where a solution is rapidly found (in a linear time) and the
search tree essentially reduces to a single branch shown
Figure~\ref{tree}. We start with some general comments on the dynamics
induced by DPLL (section~\ref{remarks}), useful to understand how 
the trajectory followed by the instance
can be computed in the $p,\alpha$ plane (section \ref{analy}). These
two first sections merely expose some previous works by Chao and
Franco, and the reader is asked to consult \cite{Fra} for more
details.  In the last section \ref{compbra}, our numerical and
analytical results for the solving complexity are presented.

In this Section, as well as in Sections~\ref{sectree} and \ref{secmixed},
the ratio of the 3-SAT instance to be solved will be denoted by $\alpha _0$.

\subsection{Remarks on the dynamics of clauses.}\label{remarks}

\subsubsection{Dynamical flows of populations of clauses.}

As pointed out in Section~\ref{deuxpp}, under the action of DPLL, some clauses
are eliminated while other ones are reduced. Let us call $C_j (T)$ the
number of clauses of length $j$ (including $j$ variables),
once $T$ variables have been assigned by the solving procedure. 
$T$ will be called hereafter ``time'', not to be confused with the computational
time necessary to solve a given instance. At time $T=0$, we obviously
have $C_3 (0)= \alpha_0 N$, $C_2 (0)=C_1 (0)=0$. As some Boolean variables are
assigned, the time $T$ increases and clauses of length one or two are
produced. A sketchy picture of DPLL dynamics at some instant $T$ is 
proposed in Figure~\ref{recip}.

We call $e_1, e_2, e_3$ and $w_2, w_1$ the flows of clauses represented 
in Figure~\ref{recip} when times increases from $T$ to $T+1$, that
is, when one more variable is chosen by DPLL after $T$ have already
been assigned. The evolution equations for the
three populations of clauses read,
\begin{eqnarray}
C_3 (T+1) &=& C_3 (T) - e_3 (T) -w_2 (T) \nonumber \\
C_2 (T+1) &=& C_2 (T) - e_2 (T) + w_2 (T) - w_1 (T) \nonumber \\
C_1 (T+1) &=& C_1 (T) - e_1 (T) + w_1 (T)  \qquad .
\label{evolsto}
\end{eqnarray}
The flows $e_j$ and $w_j$ are of course random variables that depend
on the instance under consideration at time $T$, and on the choice of
the variable (label and value) done by DPLL.  For a single descent,
{\em i.e.} in the absence of backtracking, the evolution process
(\ref{evolsto}) is Markovian and unbiased. The
distribution of instances generated by DPLL at time $T$
is uniform over the set of all the instances having $C_j (T)$ clauses
of length $j=1,2,3$ and drawn from a set of $N-T$ variables\cite{Fra}.

\subsubsection{Concentration of distributions of populations.}
\label{concent}

As a result of the additivity of (\ref{evolsto}), some concentration
phenomenon takes place in the large size limit. The numbers of clauses
of lengths 2 and 3, {\em a priori} extensive in $N$, do not
fluctuate too much,
\begin{equation}
C_j (T) = N \; c_j \left( \frac TN \right) + o (N) \qquad (j=2,3)
\ . \label{tempsreduit}
\end{equation}
where the $c_j$'s are the densities of clauses of length $j$ averaged
over the instance (quenched disorder) and the choices of variables
(``thermal'' disorder). In other words, the populations of 2- and
3-clauses are self-averaging quantities and we shall attempt at
calculating their mean densities only. Note that, in order to prevent
the occurrence of contradictions, the number of unitary clauses must
remain small and the density $c_1$ of 1-clauses has to vanish.

\subsubsection{Time scales separation and deterministic vs. stochastic
evolutions.}

Formula (\ref{tempsreduit}) also illustrates another essential feature
of the dynamics of clause populations. Two time scales are at
play. The short time scale, of the order of the unity, corresponds to
the fast variations of the numbers of clauses $C_j (T)$
($j=1,2,3$). When time increases from $T$ to $T+O(1)$ (with respect to
the size $N$), all $C_j$'s vary by $O(1)$ amounts. Consequently, the
densities of clauses $c_j$, that is, their numbers divide by $N$, are
not modified. The densities $c_j$s evolve on a long time scale of the
order of $N$ and depend on the reduced time $t=T/N$ only.

Due to the concentration phenomenon underlined above, the densities
$c_j(t)$ will evolve deterministically with the reduced time $t$. We
shall see below how Chao and Franco calculated their values. On the
short time scale, the relative numbers of clauses $D_j (T) = C_j (T) -
N c_j ( T/N)$ fluctuate (with amplitude $\ll N$) and are stochastic
variables. As said above the evolution process for these relative
numbers of clauses is Markovian and the probability rates (master
equation) are functions of slow variables only, {\em i.e.} of the
reduced time $t$ and of the densities $c_2$ and $c_3$ only. As a
consequence, on intermediary time scales, much larger than unity and
much smaller than $N$, the $D_j$s may reach some stationary
distribution that depend upon the slow variables.

This situation is best exemplified in the case $j=1$ where $c_1 (t)=0$
as long as no contradiction occurs and $D_1 (T) = C_1 (T)$.  Consider
for instance a time delay $1 \ll \Delta T \ll N$, e.g.  $\Delta T =
\sqrt N $. For times $T$ lying in between $T_0= t\, N $ and $T_1 = T_0
+\Delta T = t\, N + \sqrt N$, the numbers of 2- and 3-clauses fluctuate
but their densities are left unchanged and equal to $c_2(t)$ and $c_3
(t)$. The average number of 1-clauses fluctuates and follows some
master equation whose transition rates (from $C_1'= C_1 (T)$ to 
$C_1=C_1 (T+1)$) define a matrix ${\cal M}(C_1,C'_1)$ and depend on $t, c_2 , 
c_3$ only.  ${\cal
M}$ has generally a single eigenvector $\mu (C_1)$ with eigenvalue
unity, called equilibrium distribution, and other eigenvectors with
smaller eigenvalues (in modulus). Therefore, at time $T_1$, $C_1$ has
forgotten the ``initial condition'' $C_1 (T_0)$ and is distributed
according to the equilibrium distribution $\mu (C_1 )$ of the master
equation. 

To sum up, the dynamical evolution of the clause populations may be 
seen as a slow  and deterministic evolution of the 
clause densities to which are superimposed fast, small fluctuations. 
The equilibrium distribution of the latter adiabatically follows 
the slow trajectory.

\subsection{Mathematical analysis.}\label{analy}

In this section, we expose Chao and Franco's calculation of the
densities of 2- and 3-clauses.

\subsubsection{Differential equations for the densities of clauses.}

Consider first the evolution equation (\ref{evolsto}) for the number
of 3-clauses. This can be rewritten in terms of the average density
$c_3$ of 3-clauses and of the reduced time $t$,
\begin{equation}
\frac {d c_3 (t)} {dt}  = - z_3 (t) 
\quad ,
\end{equation}
where $z_3  = \langle e_3  + w_2  \rangle$ denotes the 
averaged total outflow of 3-clauses (Section~\ref{concent}).

At some time step $T \to T+1$, 
3-clauses are eliminated or reduced if and only if they contain the 
variable chosen by DPLL. Let us first suppose that the variable is
chosen in some 1- or 2-clauses. A 3-clause will include this variable
or its negation with probability $3/(N-T)$ and disappear with the same
probability. Due to the uncorrelation of clauses, we obtain $z_3 (t) = 
3 c_3(t)/(1-t)$. If the literal assigned by DPLL is chosen among some
3-clause, this result has to be increased by one (since this clause
will necessarily be eliminated) in the large $N$
limit. 

Let us call $\rho _j (t)$ the probability that a literal is chosen by DPLL
in a clause of length $j$ (=1,2,3). The normalization of probability
imposes that
\begin{equation}
\rho _1 (t) + \rho _2 (t) + \rho _3 (t) =1 \qquad ,
\end{equation}
at any time $t$. Extending the above discussion to 2-clauses, we
obtain 
\begin{eqnarray}
\frac {d c_3 (t)} {dt} &=& - \frac{3}{1-t} c_3 (t) - \rho _3 (t)
\nonumber \\
\frac {d c_2 (t)} {dt} &=& \frac{3}{2(1-t)} c_3 (t) - 
\frac{2}{1-t} c_2 (t) - \rho _2 (t)
\quad , \label{diff}
\end{eqnarray}

In order to solve the above set of coupled differential equations, we
need to know the probabilities $\rho _j$. As we shall see, the values
of the $\rho _j$ can directly be deduced from the heuristic of choice,
the so-called generalized unit-clause (GUC) rule exposed in 
section~\ref{heuri}.

The solutions of the differential equations (\ref{diff}) will be
expressed in terms of the fraction $p$ of 3-clauses and the ratio
$\alpha$ of clauses per variable using the identities 
\begin{equation} \label{rules}
p(t) = \frac{ c_3(t)} {c_2(t) + c_3 (t) } \quad , \qquad
\alpha (t)  = \frac {c_2(t) + c_3 (t) }{1-t}
\quad . 
\end{equation}

\subsubsection{Solution for $\alpha _0 \le 2/3$.} \label{solinf}

When DPLL is launched, 2-clauses are created with an initial flow
$\langle w_2 (0) \rangle = 3\, \alpha _0/2$. Let us suppose that $\alpha
_0 \le 2/3$, {\em i.e.} $w_2(0) \le 1$. In other words, less than one
2-clause is created each time a variable is assigned. 
Since the GUC rule compels DPLL to look for literals in the smallest
available clauses, 2-clauses are immediately removed just after
creation and do not accumulate in their recipient. Unitary clauses are
almost absent and we have
\begin{equation}
\rho _1 (t) = 0 \ ; \quad \rho _2 (t) = \frac{ 3 c_3(t)} {2 (1-t)} \ ;
\quad \rho _3 (t) = 1 - \rho _2 (t) \qquad (\alpha _0 < 2/3) \ .
\label{rhoinf}
\end{equation}
The solutions of (\ref{diff}) with the initial condition
$p(0)=1,\alpha (0) = \alpha _0$ read
\begin{eqnarray}
p(t) &=& 1 \ , \nonumber \\
\alpha (t)  &=& (\alpha _0 +2 ) (1-t ) ^{3/2} - 2 (1-t) 
\quad . 
\label{eqinf}
\end{eqnarray}
Solution (\ref{eqinf}) confirms that the instance never contains an
extensive number of 2-clauses. At some final time $t_{end}$, depending
on the initial ratio, $\alpha (t_{end} )$ vanishes: no clause is left 
and a solution is found.   

\subsubsection{Solution for $2/3 < \alpha _0 < \alpha _L$.} \label{solsup}

We now assume that $\alpha _0> 2/3$, {\em i.e.} $\langle w_2 (0)
 \rangle  > 1$. In other
words, more than one 2-clause is created each time a variable is
assigned. 2-clauses now accumulate, and give rise to 
unitary clauses. Due to the GUC prescription, in presence of 1- or
2-clauses, a literal is never chosen in a 3-clause. Thus,
\begin{equation}
\rho _1 (t) =\frac{ c_2(t)} {1-t}\ ; \quad \rho _2 (t) = 1 - \rho _1
(t)  \ ; \quad \rho _3 (t) = 0 \qquad (\alpha _0> 2/3 ) \ ,
\label{rhosup}
\end{equation}
as soon as $t>0$. The solutions of (\ref{diff}) now read
\begin{eqnarray}
p(t) &=&  \frac{ 4 \alpha _0 (1-t) ^2 }
{\alpha _0 (1-t)^2 + 3 \alpha _0 +4 \ln (1-t)} \ ,
\nonumber \\
\alpha (t)  &=& \frac{\alpha _0}4 (1-t)^2 + \frac{3 \alpha _0}4 +\ln (1-t)
\quad . \label{eqsup}
\end{eqnarray}
Solution (\ref{eqsup}) requires that the instance contains an
extensive number of 2-clauses. This is true at small times since
$p'(0) = 1/\alpha _0 -3/2 < 0$. At some time $t^* >0$, depending on
the initial ratio, $p (t ^* )$ reaches back unity: no 2-clause are
left and hypothesis (\ref{rhosup}) breaks down. DPLL has therefore
reduced the initial formula to a smaller 3-SAT instance with a ratio
$\alpha ^* = \alpha (t^*)$. It can be shown that $\alpha ^* < 2/3$. 
Thus, as the dynamical process is Markovian, the further evolution of
the instance can be calculated from Section \ref{solinf}.

\subsubsection{Trajectories in the $p,\alpha$ plane.}\label{bratra}

We show in Figure~\ref{diag} the trajectories obtained for
initial ratios $\alpha _0=0.6$, $\alpha _0=2$ and $\alpha _0=2.8$.  
When $\alpha _0 > 2/3$, the trajectory first heads
to the left (Section \ref{solsup}) and then reverses to the right
until reaching a point on the 3-SAT axis at small ratio $\alpha ^*(<2/3)$
without ever leaving the sat region. Further action of DPLL leads to a
rapid elimination of the remaining clauses and the trajectory ends up
at the right lower corner S, where a solution is achieved (section
\ref{solinf}).

As $\alpha _0$ increases up to $\alpha_L$, the trajectory gets closer and
closer to the threshold line $\alpha _C (p)$. Finally, at $\alpha _L
\simeq 3.003$, the trajectory touches the threshold curve tangentially
at point T with coordinates $(p_T=2/5,\alpha _T = 5/3)$. Note the
identity $\alpha _T = 1/(1-p_T)$. 

\subsection{Complexity.}\label{compbra}

In this section, we compute the computational complexity in the range
$0\le \alpha _0\le \alpha _L$ from the previous results.

\subsubsection{Absence of backtracking.}

The trajectories obtained in section \ref{bratra} represent the
deterministic evolution of the densities of 2- and 3-clauses when
more and more variables are assigned. Equilibrium fluctuations of
number of 1-clauses have been computed by Frieze and Suen
\cite{Fri}. The stationary distribution $\mu _t (C_1)$ of the
population of 1-clauses can be exactly computed at any time $t$. The
most important result is the probability that $C_1$ vanishes,
\begin{equation}
\mu _t (0) = 1 - \alpha (t) \big( 1 - p(t) \big) \qquad .
\end{equation}
$\mu _t (0)$ (respectively $1-\mu _t (0)$)
may be interpreted as the probability that a variable 
assigned by DPLL at time $t$ is chosen
through splitting (resp. unit--propagation). 
When DPLL starts solving a 3-SAT
instance, $\mu _{t=0} (0)=1$ and many splits are necessary. If the
initial ratio $\alpha _0$ is smaller than 2/3, this statement remains
true till the final time $t_{end}$ and the absence of 1-clauses
prevents the onset of contradiction. Conversely, if $2/3 <
\alpha _0< \alpha _L$, as $t$ grows, $\mu _t (0)$ decreases are more and
more variables are fixed through unit-propagation. The population of
1-clauses remains finite and the probability that a contradiction
occurs when a new variable is assigned is $O(1/N)$ only. However small
is this probability, $O(N)$ variables are fixed along a complete
trajectory. The resulting probability that a contradiction never
occurs is strictly smaller than unity\cite{Fri},
\begin{equation}
{\cal P}_{No\ Contradiction} = \exp \left( - \frac 14
\int _{0} ^{t_{end}} \frac{dt} {1-t} \frac{ (1 - \mu _t (0))^2} { 
\mu _t (0) } \right) \qquad .
\end{equation}
Frieze and Suen have shown that contradictions have no dramatic
consequences. The number of backtrackings necessary to find a solution
is bounded from above by a power of $\log N$. The final
trajectory in the $p,\alpha$ plane is identical to the one shown in
section~\ref{bratra} and the increase of complexity is negligible with
respect to $O(N)$. 

When $\alpha _0$ reaches $\alpha _L$, the trajectory intersects the
$\alpha = 1/(1-p)$ line in T. At this point, $\mu (0)$ vanishes and
backtracking enters massively into play, signaling the cross-over to
exponential regime. 

\subsubsection{Length of trajectories.} \label{length}

From the above discussion, it appears that a solution is found by DPLL
essentially at the end of a single descent (Figure~\ref{tree}A). 
Complexity thus scales
linearly with $N$ with a proportionality coefficient $\gamma (\alpha _0)$
smaller than unity.

For $\alpha _0< 2/3$, clauses of length unity are never created by DPLL
(Section~\ref{solinf}). Thus, DPLL assigns the overwhelming
majority of variables by splittings. $\gamma (\alpha _0)$ simply equals
the total fraction $t_{end}$  of variables chosen by DPLL. From
(\ref{eqinf}), we obtain
\begin{equation} \label{const}
\gamma (\alpha _0 ) = 1 - \frac 4{(\alpha _0+2 )^2} \qquad (\alpha _0 
\le 2/3 ) \ .
\end{equation}
For larger ratios, {\em i.e.} $\alpha _0> 2/3$, the trajectory must be
decomposed into two successive portions (Section~\ref{solsup}).
During the first portion, for times $0<t<t^*$, 2-clauses are present with
a non vanishing density $c_2 (t)$. Some of these 2-clauses are reduced
to 1-clauses that have to be eliminated next. Consequently, when DPLL
assigns an infinitesimal fraction $dt$ of variables, a fraction $\rho _1 (t) = 
\alpha (t) ( 1 -p(t)) dt$ are fixed by unit-propagation only. The number of
nodes (divided by $N$) along the first part of the branch thus reads,
\begin{equation}
\gamma _1 = t^* - \int _0 ^{t^*} dt \;\alpha (t) ( 1 -p(t)) \quad .
\label{gam1}
\end{equation}
At time $t^*$, the trajectory touches back the 3-SAT axis $p=1$ at
ratio $\alpha ^* \equiv \alpha (t^*) < 2/3$. The initial instance is 
then reduces to a smaller and smaller 3-SAT formula, with a  ratio
$\alpha (t)$ vanishing at $t_{end}$. According to the above discussion, the
length of this second part of the trajectory equals
\begin{equation}
\gamma _2 = t_{end} - t^* \quad .
\label{gam2}
\end{equation}
It results convenient to plot the total complexity $\gamma = \gamma
_1 + \gamma _2$ in a parametric way. To do so, we express the initial
ratio $\alpha _0$ and the complexity $\gamma$ in terms of the end time 
$t^*$ of the first part of the branch. A simple calculation from
equations (\ref{eqsup}) leads to
\begin{eqnarray}
\alpha _0 (t^* ) &=& - \frac {4 \ln (1-t^*)}{3 t^* (2 - t^* )} \nonumber
\\ \gamma (t^* ) &=& 1 - \frac{4 (1-t^*)}{( 2 + (1-t^*)^2 \alpha _0 (t^*)
)^2} +t^* + (1-t^*) \ln (1-t^*)  - \frac 14 \alpha _0 (t^*) \; (t^*
)^2 (3 -t^*)
\label{constparam}
\end{eqnarray}
As $t^*$ grows from zero to $t^* _L \simeq 0.892$, the initial ratio
$\alpha _0$ spans the range $[2/3;\alpha _L]$. 
The complexity coefficient $\gamma (\alpha _0)$ can be computed 
from equations (\ref{const},\ref{constparam}) with the results  
shown Figure~\ref{constante}. 
The agreement with the numerical data of Section~\ref{secnumlin} is excellent.

\section{Tree trajectories and the exponential regime (unsat phase).}
\label{sectree}

To present our analytical study of the exponentially large search
trees generated when solving hard instances, we consider first a
simplified growth tree dynamics in which variables, on each branch,
are chosen independently of the 1- or 2-clauses and all branches split
at each depth $T$. This toy model is too simple a growth process to
reproduce a search tree analogous to the ones generated by DPLL on
unsat instances. In particular, it lacks two essential ingredients of
the DPLL procedure: the generalized unit clause rule 
(literals are chosen from the shortest clauses), and the possible
emergence of contradictions halting a branch growth.  Yet, the study
on the toy model allows us to expose and test the main analytical 
ideas, before turning to the full analysis of DPLL in Section~\ref{sectree2}.

\subsection{Analytical approach for exponentially large search trees: 
a toy case.}

\label{seces}

\subsubsection{The toy growth dynamics.}

In the toy model of search tree considered hereafter, only 3-clauses
matter. Initially, the search tree is empty and the 3-SAT instance
is a collection of $C'_3$ 3-clauses drawn from $N$ variables. Next,
a variable is randomly picked up and fixed to true or false, leading
to the creation of one node and two outgoing branches carrying
formulae with $C_3$ and $\hat C_3$ 3-clauses respectively, see
Figure~\ref{algotoy}. This elementary process is then repeated for
each branch. At depth or ``time'' $T$, that is when $T$ variables have been
assigned along each branch, there are $2^T$ branches
in the tree.

The quantity we focus on is the number of branches having
a given number of 3-clauses at depth $T.$ On each branch,
after  a variable has been assigned, $C_3$ decreases (by clause reduction 
and elimination) or remains unchanged (when the chosen variable 
does not appear in the clauses).  So, the recipients 
of 3-clauses attached to each branch, see  Figure~\ref{recip}, 
leak out with time T.

We now assume that each branch in the tree evolves independently of
the other ones and obeys a Markovian stochastic process. This amounts
to neglect correlations between the branches, which
could arise from the selection of the same variable on
different branches at different times. As a consequence of 
this decorrelation approximation, the value of $C_3$ at time $T+1$
is a random variable whose distribution depends only on $C'_3$ (and time 
$T$).

The decorrelation approximation allows to write a master equation
for the average number $B(C_3;T)$ of branches carrying $C_3$ 3-clauses
at depth $T$ in the tree,
\begin{equation}
\label{bra}
B({ C_3};T+1)=\sum_{C_3'=0}^{\infty}\;K\;({ C_3},{ C_3'};T)\;
B({C'_3};T) \qquad .
\end{equation}
$K$ is the branching
matrix; the entry $K(C_3,C'_3)$ is the averaged number of branches 
with ${ C_3}$ clauses issued
from a branch carrying $C_3'$ clauses after one variable is
assigned. This number lies between zero (if $C_3>C_3'$) and two (the
maximum number of branches produced by a split), and is easily deduced 
from the evolution of the recipient of 3-clauses in Figure~\ref{recip},
\begin{equation}
\label{bin}
 K\;({ C_3},{ C_3'};T)=2\;\chi\,(C_3'-C_3)\,\left( \begin{array} {c}
 C_3' \\ C_3'-C_3
\end{array} \right) \; \left( \frac 3{N-T}
\right)^{C_3'-C_3} \;\left (1-\frac 3{N-T} \right)^{C_3}
\end{equation}
The factor 2 comes from the two branches created at each split;
$\chi (C_3'-C_3)$ equals unity if $C_3'-C_3 \ge 0$, and zero otherwise.  The
binomial distribution in (\ref{bin}) comes from the 
probability that the variable fixed at
time $T$ appears exactly in $C_3'-C_3$ 3-clauses.

\subsubsection{Partial differential equation for the distribution of branches.}
\label{sec56}

For large instances $N,M \to \infty $, this binomial distribution 
simplifies to a Poisson distribution, with parameter
$m _3(T)=3\;C_3'/(N-T)$. The branching matrix (\ref{bin}) thus reads,
\begin{equation}
\label{pois}
K\;({ C_3},{ C_3'};T) \simeq
K\;({ C_3'-C_3}, m _3(T)) = 2\;\chi\,(C_3'-C_3)\; e^{-m _3(T)} \; \frac{
m _3(T)^{C_3'-C_3}}{\;(C_3'-C_3)!}
\end{equation}
Consider now the variations of the entries of $K$ (\ref{pois}) over a time 
interval $T_0 = t\, N \le T \le T_1 = (t+\epsilon )\, N$. 
Here, $\epsilon$ is a small
parameter but of the order of one with respect to $N$. In other words,
we shall first send $N$ to infinity, and then $\epsilon$ to zero.
$m_3(T)$ weakly varies between $T_0$ and $T_1$: 
$m_3 (t)=3\;c_3'/(1-t)+O(\epsilon)$ where $c_3'=C_3'/ N$ is the intensive
variable for the number of 3-clauses. The branching matrix (\ref{pois}) 
can thus be rewritten, for all times $T$ ranging from $T_0$ to $T_1$, as
\begin{equation}
\label{pois2}
K(C_3'-C_3,m _3(t)) = 2\;\chi\,(C_3'-C_3)\; e^{-m_3(t)} \; \frac{
m_3(t)^{C_3'-C_3}}{\;(C_3'-C_3)!} + O(\epsilon) \qquad ,
\end{equation}
We may now iterate eqn.(\ref{bra}) 
over the whole time interval of length ${\cal T} = T_1 -T_0=
\epsilon \, N$,
\begin{equation}
 \label{bra2} B({ C_3}; t\; N +{\cal T})=\sum_{C'_3=0}^{\infty}\;
\left[K ^{\cal T}\right] \;({ C_3'-C_3};t) \; B({C_3'}; t\; N) \qquad ,
\end{equation}
where $K^{\cal T}$ denotes the ${\cal T}^{th}$ power of $K$. 
As $K$ depends only on the difference $C_3'-C_3$,
it can be diagonalized by plane waves $v(q _3,C_3)=e^{i\, q_3\, 
C_3}/ \sqrt{2\pi}$ 
with wave numbers $0\le q_3 <2\,\pi $ (because $C_3$ is an integer-valued 
number). The corresponding eigenvalues read
\begin{equation}
\lambda _{q_3} (t)= 2\;\exp\bigg[ m _3(t)\,(e^{i\,q _3}-1) \bigg]
\end{equation} 
Reexpressing matrix (\ref{pois2}) using its eigenvectors and
eigenvalues, equation (\ref{bra2}) reads
\begin{equation}
\label{bra3}
B({ C_3}; t\, N+{\cal T})=\sum_{C'_3=0}^{\infty}\;\int_0^{2\pi}\, \frac{
dq _3}{2\pi} \, \big( \lambda_{q _3} (t) \big) 
^{\cal T}\; e^{i(C_3-C'_3)\,q}\; B({C_3'};t \, N)
\end{equation}
Branches duplicates at each time step and proliferate exponentially with
$T =t\, N$. A sensible scaling behavior for their number is, for any
fixed fraction $c_3$,
\begin{equation}
\label{scal}
\lim_{N\to\infty}\; \frac 1N \ln B(c_3\, N; t\, N)= \omega(c_3,t)
\end{equation}
Note that $\omega$ has to be divided by $\ln 2$ to obtain
the logarithm of the number of branches in base 2 to match the
definition of Section~\ref{foc}. Similarly, we introduce the 
rescaled variable $r_3$, replacing $C_3'$ in the sum of (\ref{bra2}),
\begin{equation}
\label{tempo}
C'_3-C_3=r_3\, \cal{T} \qquad .
\end{equation}
Equation (\ref{tempo}) simply means that $r_3$ is of the order of unity 
when ${\cal T}$ gets very large, since the number of 3-clauses that 
disappear after having fixed ${\cal T}$ variables is typically of the
order of ${\cal T}$. 
We finally obtain from eqn.(\ref{bra3}),
\begin{equation}
\label{bra4}
\exp \bigg( N\;\omega(c_3,t+\epsilon) \bigg)
=\frac 1{\cal T}\;\int_{0}^{\infty} dr_3
\;\int_0^{2\pi}\; \frac{dq _3}{2\pi}
\; \exp \bigg[  N\; \big( \epsilon \, \ln \lambda _{q_3} (t) -i\, \epsilon
\,r _3\,q_3 + \omega(c_3+\epsilon\,r_3 , t ) \big) \bigg] \qquad .
\end{equation}
In the $N\to \infty$ limit, the integrals in (\ref{bra4}) may be
evaluated using the saddle point method,
\begin{equation}
\label{bra47}
\omega(c_3,t+\epsilon) = \max_{r _3,q _3}
\bigg[ \epsilon \, \ln \lambda _{q_3} (t) -i\, \epsilon
\,r _3\,q_3 + \omega(c_3+\epsilon\,r_3 , t )  \bigg] + O(\epsilon ^2)
\qquad ,
\end{equation}
due to the terms neglected in (\ref{pois2}).
Assuming that $\omega (c_3, t)$ is a partially
differentiable functions of its arguments, and expanding (\ref{bra47})
at the first order in $\epsilon$, we obtain the 
partial differential equation,
\begin{equation}
\label{par}
\frac {\partial \omega}{\partial t}(c_3,t)= \max_{r _3,q_3}\left (\ln
\lambda _{q_3} (t) -i\,r_3 \, q_3 +r_3\,\frac{\partial \omega}{\partial
c_3}(c_3,t)\right)
\end{equation}
The saddle point lies in $q _3=-i\frac {\partial \omega}{\partial
c_3}$, leading to,
\begin{equation}
\label{par2}
\frac {\partial\omega} {\partial t} (c_3,t)= \ln 2 -
\frac {3\,c_3}{1-t} + 
\frac {3\,c_3}{1-t} \;\exp \left[ \frac {\partial\omega} {\partial
c_3} (c_3,t) \right] \qquad .
\end{equation}
It is particularly interesting to note that a partial differential
equation emerges in (\ref{par2}). In contradistinction
with the evolution of a single branch described by a set of ordinary
differential equations (Section~\ref{secbranch}), the analysis
of a full tree requires to handle information about the whole
distribution $\omega$ of branches, and so to a partial
differential equation. 

\subsubsection{Legendre transform and  solution of the partial 
differential equation.}
\label{sopde}

To solve this equation is convenient to define the Legendre
transformation of the function $\omega(c_3,t)$,
\begin{equation}
\label{legendre}
\varphi (y _3,t)=\max_{c_3}\, \bigg[ \omega(c_3,t)+y_3\,c_3 \bigg]
\end{equation}
From a statistical physics point of view this is equivalent to pass,
at fixed time $t$, from a microcanonical `entropy'
$\omega$ defined as a function of the `internal energy' $c_3$, to a `free
energy' $\varphi$ defined as a function of the `temperature' $y_3$.  More
precisely, $\varphi(y_3,t)$ is the logarithm divided by $N$ of the generating
function of the number of nodes $B(C_3;t\;N)$. Equation
(\ref{legendre}) defines the Legendre relations between $c_3$ and
$y$,
\begin{equation}
\label{con1}
\left .\frac {\partial \omega}{ \partial {c_3}}\right |_{c_3(y _3)}=-y_3
\qquad \hbox{\rm and}\qquad 
\left .\frac{ \partial \varphi} {\partial {y _3}}
 \right |_{y_3 (c _3)} =c_3 \quad .
\end{equation}
In terms of $\varphi$ and $y_3$, the partial differential equation
(\ref{par2}) reads
\begin{equation}
\label{difz}
\frac{\partial \varphi}{\partial t}(y_3,t)=\ln 2- \frac 3 {1-t} \;(1-e^{-y_3}) 
\;\frac{\partial \varphi}{\partial y_3} (y_3,t) \qquad ,
\end{equation}
and is linear in the partial derivatives. This is a consequence of
the Poissonian nature of the distribution entering $K$ (\ref{pois}).  
The initial condition for the
function $\varphi (y_3,t)$ is smoother than for $\omega(c_3,t)$. At time $t=0$,
the search tree is empty: $\omega (c_3,t=0)$ equals zero for 
$c_3=\alpha_0$, and $- \infty$ for $c_3 \ne \alpha_0$. The Legendre
transform is thus $\varphi (y_3,0)= \alpha_0 \; y_3$, a linear function of 
$y_3$. The solution of eqn.(\ref{difz}) reads
\begin{equation} 
\varphi(y _3,t)=t\;\ln 2+\alpha_0
\ln\,\left[1-\left(1-e^{y _3}\right)\,\left(1-t\right)^3\right]
\end{equation}
and, through Legendre inversion,
\begin{equation}
\omega(c_3,t)=t\,\ln 2+3c_3\,\ln(1-t)+\alpha_0 \ln\alpha_0-c_3\ln c_3
-(\alpha_0-c_3) \ln \left[\frac{\alpha_0 -c_3}{1-(1-t)^3} \right]
\qquad ,
\end{equation} 
for $0<c_3<\alpha _0$, and $\omega=-\infty$ outside this range.
We show in Figure~\ref{entroy} the behavior of $\omega (c_3,t)$ for
increasing times $t$. The curve has a smooth and bell-like shape, with a 
maximum located in $\hat c_3 (t) = \alpha _0 (1-t)^3 ,
\hat \omega (t)= t \ln 2$. The number of branches at depth $t$ equals
$B(t) = e^{N\, \hat \omega (t)} = 2^{N \,t}$,
and almost all branches carry  $\hat c_3(t)\, N$ 3-clauses,
in agreement with the expression for $c_3 (t)$ found for the simple 
branch trajectories in the case $2/3<\alpha_0<3.003$ (Section \ref{solsup}). 
The top of the curve $\omega (c_3,t)$, at fixed $t$, provides direct 
information about the dominant, most numerous branches.

For the real DPLL dynamics studied in next section, 
the partial differential equation for the growth process is much more
complicated, and exact analytical solutions are not available.
If we focus on exponentially dominant branches, we may 
as a first approximation follow the dynamical evolution of the points
of the curve $\omega(c_3,t)$ around the top.  To do so, we linearize 
the partial differential equation (\ref{difz}) for the Legendre transform
$\phi$ around the origin $y_3=0$ (\ref{con1}), 
\begin{equation}
\frac{\partial \varphi}{\partial t}(y _3,t) \simeq \ln 2- \frac 3 
{1-t} \; y_3 \;\frac{\partial \varphi}{\partial y _3} (y_3,t) \qquad .
\end{equation}
The solution of the linearized equation,
$\varphi(y _3,t)=t\;\ln 2 +\alpha_0 (1-t)^3\; y_3$,
is itself a linear function of $y_3$. Through Legendre
inversion, the slope gives us the coordinate $\hat c_3 (t)$ 
of the maximum of $\omega$,
and the constant term, the height $\hat \omega (t) $ of the top.

%%%%%%%%%%%%%%%%%%%%%%%%%%%%%%%%%%%%%%%%%%%%%%%%%%%%%%%%%%%%%%

\subsection{Analysis of the full DPLL dynamics in the unsat phase.}

\label{sectree2}

\subsubsection{Parallel versus sequential growth of the search tree.}
\label{paral}

A generic refutation search tree in the unsat phase is shown in 
Figure~\ref{tree}B. It is the output of a {\em sequential} 
building process: nodes and edges are added by DPLL through successive 
descents and backtrackings.  We have imagined a different building up, 
that results in the same complete tree but can be mathematically 
analyzed: the tree grows in {\em parallel}, layer after layer. A new 
layer is added by assigning, according to DPLL heuristic, one more 
variable along each living branch. As a result, some branches split, 
others keep growing and the remaining ones carry contradictions and die out.

\subsubsection{Branching matrix for DPLL.}

To take into account the operation of the DPLL procedure, see 
Section~\ref{secdpll}, we follow, in each branch, the number of 3-clauses 
$C_3$ as well as the numbers $C_2$ and $C_1$ of 2- and 1-clauses.
The evolution equation for the average number of branches $B$ carrying
instances with $C_j$ $j$-clauses ($j=1,2,3$) reads, from (\ref{bra}),
\begin{equation}
\label{bradp}
B({C_1,C_2,C_3};T+1)=\sum_{C_1',C_2',C_3'=0}^{\infty}\;K\;({C_1,C_2,
C_3},{C_1',C_2', C_3'};T)\; B({C_1',C_2',C'_3};T) \qquad ,
\end{equation}
where the branching matrix $K$ now equals
\begin{eqnarray}
\label{bbra}
&& K (C_1,C_2,C_3; C_1',C_2',C_3'; T) = {C_3' \choose C_3'-C_3} \; \left
(\frac 3{N-T}\right)^{C_3'-C_3} \; \left(1-\frac 3{N-T}\right)^{C_3}
\sum_{w_2=0}^{C_3'-C_3} \left (\frac
1 2 \right)^{C_3'-C_3} {C_3'-C_3\choose w_2} \times
\nonumber \\ 
&&  \left \{(1 - \delta _{C'_1} )\;
\sum_{z_2=0}^{C_2'} {C_2' \choose z_2} \left (\frac
2{N-T}\right)^{z_2}\, \left (1- \frac 2{N-T}\right)^{C_2'-z_2} \right.
\sum_{w_1=0}^{z_2} \left (\frac 1 2
\right)^{z_2} {z_2\choose w_1}\; \delta_{C_2-C_2'-w_2+z_2}
\;\delta_{C_1-C_1'-w_1+1}  + \delta_{C_1'} \times \nonumber \\ && \left. 
\sum_{z_2=0}^{C_2'-1} {C_2'-1 \choose z_2} \left (\frac
2{N-T}\right)^{z_2}\, \left( 1- \frac 2{N-T}\right)^{C_2'-1-z_2}
\sum_{w_1=0}^{z_2} \left(\frac 1 2 \right)^{z_2} {z_2\choose w_1}\;
\delta_{C_2-C_2'-w_2+z_2+1} \; [\delta_{C_1-w_1} + \delta_{C_1-1-w_1}]
 \right\} \ ,
\end{eqnarray}
where $\delta _C$ denotes the Kronecker delta function:
$\delta _C=1$ if $C=0$, $\delta _C=0$ otherwise.

To understand formula (\ref{bbra}), the picture of recipients in
Figure~\ref{recip} proves to be useful.  $K$ expresses the average
flow of clauses into the sink (elimination), or to the recipient to the
right (reduction), when passing from depth $T$ to depth $T+1$ by fixing
one variable. Among the $C_3'-C_3$ clauses that flow out from the
leftmost 3-clauses recipient, $w_2$ clauses are reduced and go into
the 2-clauses container, while the remaining $C_3'-C_3-w_2$ are
eliminated.  $w_2$ is a random variable in the range $0 \le w_2
\le C_3'-C_3$ and drawn from a binomial distribution of parameter $1/2$, which
represents the probability that the chosen literal is the negation of the
one in the clause.

We have assumed that the algorithm never chooses the variable among
3-clauses. This hypothesis is justified {\it a posteriori} because in
the unsat region, there is always (except at the initial time $t=0$)
an extensive number of 2-clauses.  Variable are chosen among 1-clauses,
or if none is present, among 2-clauses. The term on the r.h.s. of
eqn. (\ref{bbra}) beginning with
$\delta_{C_1}$ (respectively $1-\delta_{C_1}$) corresponds to the latter
(resp. former) case. 
$z_2$ is the number of clauses (other than the one from which
the variable is chosen) flowing out from the second recipient; it
obeys a binomial distribution with parameter $2/(N-T)$, equal to the
probability that the chosen variable appears in a 2-clause. The
2-clauses container is, at the same time, poured with $w_2$
clauses. In an analogous way, the unitary clauses recipient welcomes
$w_1$ new clauses if it was empty at the previous step. If not, a
1-clause is eliminated by fixing the corresponding literal.

The branch keeps growing as long as the level
$C_1$ of the unit clauses recipient remains low, {\em i.e.} $C_1$ 
remains of the order of unity so that the probability to have two, or 
more, 1-clauses with opposite literals can be neglected.  For
this reason, we do not take into account the (extremely rare) event
of 1-clauses including equal or opposite literals
and $z_1$ is always equal to zero.  We shall consider later a halt 
criterion for the tree growth process, see dot-dashed line in the phase 
diagram of Figure~\ref{diag}, where the condition $C_1 =O(1)$ 
breaks down due to an avalanche of 1-clauses.  

Finally, we sum over all possible flow values $w_2,z_2,w_1$ that 
satisfy the conservation laws $C_2-C_2'=w_2-z_2$, $C_1-C_1'=w_1-1$ when
$C_1'\ne 0$ or, when $C'_1=0$,  $C_2-C_2'=w_2-z_2-1$, $C_1=w_1$ if
the literal is the same as the one in the clause or $C_1=w_1+1$ if the literal
is the negation of the one in the clause. The presence of two $\delta$ is
responsible for the growth of the number of branches.  In the real
sequential DPLL dynamics, the inversion of a literal at a
node requires backtracking; here, the two edges grow in
parallel at each node according to Section~\ref{paral}.
 
%Note that the above conservation laws correspond to the evolution equations
%(\ref{evolsto}) upon writing $e_3+ w_2=z_3, e_2+w_1-(1-\delta _{C_1})=z_2,
%e_1=\delta _{C_1}$, $z_1=0$ and considering also the possibility of
%literal inversion $w_1 \to w_1+1$.

In the large $N$ limit, the matrix (\ref{bbra}) can be written
 in terms of Poisson distributions through the introduction of the
parameters $m_3=3\, C_3 /(N-T) $ and $m_2=
2\, C_2 / (N-T) $, see eqn. (\ref{pois}).

\subsubsection{Ground state of the branching matrix and localization 
properties.} 

Due to the translational invariances of $K$ in $C'_3 -C_3$ and
$C'_2-C_2$, the vectors
\begin{equation}
\label{eigenve}
v _{q,q_2,q_3}(C_1',C_2',C_3')=e^{i(q_2 C_2'+q_3 C_3')} \,\tilde{v}_q(C_1)
\end{equation}
are eigenvectors of the matrix $K$ with
eigenvalues
\begin{equation}
\label{eigenva1}
\lambda_{q,q_2,q_3}= \exp \left\{m_3 \left [ \frac{e^{iq_3}} 2 \left
(1+e^{-iq_2} \right) -1 \right]\right\} \tilde{\lambda}_{q,q_2}
\qquad ,
\end{equation}
if and only if $\tilde{v} _q(C_1)$ is an eigenvector, with 
eigenvalue $\tilde{\lambda}_{q,q_2}$, of the reduced matrix
\begin{equation}
\label{redmat}
\tilde K _{q_2}(C_1,C_1')= \sum_{z_2=0}^{\infty} e^{-m_2}\,
\frac{( m_2\; e^{iq_2}) ^{z_2}}{z_2!}\;\sum_{w_1=0}^{z_2} {z_2\choose w_1} 
\frac{1}{2^{z_2}}
\bigg\{\delta_{C_1-C_1'-w_1+1}\; (1-\delta_{C_1'})+
\,e^{i\,q_2}(\delta_{C_1-w_1}+\delta_{C_1-1-w_1})\;\delta_{C_1'}\bigg\}
\ .
\end{equation}
Note that, while $q_2,q_3$ are wave numbers, $q$ is a formal index
used to label eigenvectors. The matrix $\tilde K _{q_2}$ (\ref{redmat}) has
been obtained by applying $K$ onto the vector $v$ (\ref{eigenve}), and
summing over $C_3', C_2', w_2$ and $z_3$. 

The diagonalization of the non hermitian matrix $\tilde K_{q_2}$ is
exposed in Appendix~\ref{diagosing}, and relies upon the introduction of the
generating functions of the eigenvectors $\tilde{v} _q$,
\begin{equation}
\tilde V_q(x)=\sum_{C_1=0}^{\infty}\tilde{v_q}(C_1)\;x^{C_1}
\qquad .
\end{equation}
The eigenvalue equation for $\tilde K_{q_2}$ translates into a 
self-consistent equation for  $\tilde V _q (x)$, the singularities
of which can be analyzed in the $x$ plane, and permit to calculate the largest 
eigenvalue\footnote{We show in next Section that $q_2$ is purely imaginary
at the saddle-point, and therefore the eigenvalue in eqn.(\ref{eiggf2}) is
real-valued. } of $\tilde K_{q_2}$,
\begin{equation}
\label{eiggf2}
\tilde \lambda_{q_2}= \frac{2}{-1+\sqrt{1+4\,e^{-iq_2}}}
\; \exp\left\{ - m_2 + m_2 \, \frac{e^{i\,q_2}} 4 \left
(1+\sqrt{1+4\,e^{-iq_2}}\right) \right\}
\end{equation}   
The properties of the corresponding, maximal eigenvector $\tilde v_0$ 
are important.
Depending on parameters $q_2$ and $m_2$, $\tilde{v}_0(C_1)$ is either 
localized around small integer values of $C_1$ (the average 
number of 1-clauses is of the order of the unity), or extended (the
average value of $C_1$ is of the order of $N$). As contradictions inevitably 
arise when $C_1= O(N)$, the delocalization transition undergone by the
maximal eigenvector provides a halt criterion for the growth of the tree.

\subsubsection{Partial differential equation for the search tree growth.}
\label{sec57}

Following Section~\ref{sec56} and eqn.(\ref{bra3}), we write the
evolution of the number of branches  between times $t\, N$
and $t\, N +{\cal T}$ using the spectral decomposition of  $K$, 
\begin{eqnarray}
\label{evolt}
B(C_1,C_2,C_3 ; t\,N+{\cal T})= \sum_{C_1',C_2',C_3'} \sum_q
\int_0^{2\pi} \,\frac{dq_2}{2 \pi} \,\frac{dq_3}{2\pi}
&&\; e^{i[q_3\,(C_3-C_3')+q_2\,(C_2-C_2')]}
\, \tilde{v}_q(C_1)\;
\tilde{v}_q^+(C_1')\; \times \nonumber \\
&& ( \lambda_{q,q_2,q_3} )^{\cal T}\;B(C'_1,C'_2,C'_3,t\,N) \qquad .
\end{eqnarray}
$\sum_q$ denotes the (discrete or continuous) sum on all eigenvectors
and $\tilde{v}_q^+$ the left eigenvector of $\tilde K$.  We make
the adiabatic hypothesis that the probability to have $C_1$  unit clauses
at time $t\,N+{\cal T}$ becomes stationary on the time scale $1\ll 
{\cal T} \ll N$ and is independent of the number $C_1'$ of 1-clauses
at time $t\, N$  (Section~\ref{secbranch}).  
As ${\cal T}$ gets large, and at fixed $q_2,q_3$,
the sum over $q$ is more and more dominated by the
largest eigenvalue $q=0$, due to the gap between the first
eigenvalue (associated to a localized eigenvector) and the continuous 
spectrum of delocalized eigenvectors (Appendix~\ref{diagosing}).  
Let us call $\Lambda (q_2,q_3) \equiv \lambda _{0,q_2,q_3}$ this 
largest eigenvalue, obtained from
equations (\ref{eiggf2}) and (\ref{eigenva1}).
Defining the average of the number of branches over the equilibrium
distribution of 1-clauses,
\begin{equation}
B(C_2,C_3;T)=\sum_{C_1} B(C_1,C_2,C_3) \; \tilde{v}_0 ^\dagger (C_1) \qquad ,
\end{equation}
equation (\ref{evolt}) leads to
\begin{equation}
B(C_2,C_3 ; t\,N+{\cal T})=\sum_{C_2',C_3'=0}^\infty \,\int_0^{2\pi}\,
\frac{dq_2}{2\pi} \,\frac{dq_3}{2 \pi}  \;e^{i[q_3\,(C_3-C_3')+
q_2\,(C_2-C_2')]} \big(\Lambda (q_2,q_3) \big)^{\cal T} \;B(C'_2,C'_3 ;t\,N)
\quad . \label{tgb}
\end{equation}
The calculation now follows closely the lines of Section~\ref{sec56}.
We call $\omega(c_2,c_3,t) = \lim _{N\to\infty}\; \ln B(c_2\,N,c_3\,N
;t\,N) /N$, the logarithm of the number of branches carrying an instance
with $c_2\,N$ 2-clauses and $c_3\,N$ 3-clauses at depth $t\,N$.
Similarly, we rewrite the sums on $C'_2,C'_3$ on the r.h.s. of
eqn.(\ref{tgb}) as integrals over the reduced variable 
$r_2=(C_2'-C_2)/{\cal T}$, $r_3=(C_3'-C_3)/{\cal T}$, 
see equations~(\ref{bra3}) and (\ref{bra4}).
A saddle-point calculation of the four integrals over $q_2,q_3,r_2,r_3$
can be carried out, resulting in a partial differential equation for
$\omega (c_2,c_3,t)$, 
\begin{equation}
\label{parf}
\frac {\partial \omega}{\partial t}(c_2,c_3,t)= \ln \Lambda \left(
-i\frac {\partial \omega}{\partial
c_2}, -i\frac {\partial \omega}{\partial c_3} \right)
\qquad ,
\end{equation}
or, equivalently,
\begin{equation}
\frac {\partial \omega}{\partial t}=\frac {\partial \omega}{\partial
c_2}+ \ln \left[ \frac{1+\sqrt{4\,e^{-\frac{\partial \omega}{\partial
c_2}}+1}}2 \right]+ \frac{3\,c_3}{1-t} \left[
\frac{e^{\frac{\partial \omega}{\partial c_3}}}2\,
(1+e^{-\frac{\partial \omega}{\partial c_2}})-1 \right]+
\frac{2 c_2}{1-t} \,\left[ \frac{e^{\frac{\partial \omega}{\partial c_2}}}4
\left(1+\sqrt{4\,e^{-\frac{\partial \omega}{\partial c_2}}+1}\right)-1 \right]
\ .
\end{equation}

\subsubsection{Approximate solution of the partial differential equation}
\label{yhn}

As in Section~\ref{sopde}, we introduce the Legendre transformation
of $\omega(c_2,c_3,t)$,
\begin{equation} \label{vp}
\varphi(y_2,y_3,t)=\max_{c_2,c_3}\,\bigg(
\omega(c_2,c_3,t)+y_2\,c_2+y_3\,c_3 \bigg) \qquad .
\end{equation}
The resulting partial differential equation on $\varphi$ is given
in Appendix~\ref{quadapp}, and cannot be solved analytically. We therefore
limit ourselves to the neighborhood of the top of the surface
$\omega(c_2,c_3,t)$ through a linearization around
$y_2=y_3=0$,  
\begin{equation}
\frac {\partial \varphi}{\partial t} \simeq \ln
\left(\frac{1+\sqrt{5}}{2}\right) -\left(\frac{5+\sqrt{5}}{10}\right)
y_2- \frac {3} {1-t}\; \left(y_3-\frac{y_2}{2}\right)  \frac {\partial 
\varphi}{\partial y_3} -\frac 1{1-t} 
\left(\frac{3-\sqrt{5}}{2}+\frac{5+3\sqrt{5}}{10} y_2 \right)\frac{\partial
\varphi}{\partial y_2} \label{lipde2}
\end{equation} 
The solution of eqn (\ref{lipde2}) is given by the combination of a
particular solution with $\frac{\partial \varphi}{\partial y_3}=0$ and
the solution of the homogeneous counterpart of equation (\ref{lipde2}).
We write below the general solution for any 2+p-SAT unsat problem
with parameters $p_0 , \alpha _0 > \alpha _C(p_0)$, the 3-SAT
being recovered when $p_0=1$. The initial condition at time $t=0$ reads 
$\varphi(y_2,y_3,0)=\alpha_0\,p_0\,y_3+\alpha_0\,(1-p_0)\,y_2$. 
We obtain
\begin{equation}
\varphi(y_2,y_3,t)= \hat {c}_2  (t)\,y_2+ {\hat c}_3 (t)\,y_3+
\hat {\omega} (t) \label{solulin89}
\end{equation}
with
\begin{eqnarray}
\label{c3c}
\hat c_2 (t)&=& \alpha_0\,p_0\;\left(\frac{9\sqrt{5}+75}{116}\right)
\left[\,(1-t)^{\frac{5+3\sqrt{5}}{10}} -(1-t)^3\right] + 
\left (\alpha_0(1-p_0)+2+\sqrt{5}\right)\,(1-t)^{\frac{5+3\sqrt{5}}
{10}} -(2+\sqrt{5})(1-t) \ , \nonumber \\
\hat {c}_3(t) &=& \alpha_0\,p_0 (1-t)^3 \ , \nonumber \\
\hat \omega (t) &=&\alpha_0\,p_0 \left(\frac{15-4\sqrt{5}}{58}\right)
\,\left(1-(1-t)^3\right)+ t\left(\ln\left(\frac{1+\sqrt{5}}{2}\right)
+ \frac{1+\sqrt{5}}{2}\right) + \nonumber \\
&&\frac{7\sqrt{5}-15}{2}\,\left(\alpha_0\,p_0
\left(\frac{9\sqrt{5}+75}{116}\right)+
\alpha_0\,(1-p_0)+2+\sqrt{5}\right)\left((1-t)^{\frac{5+3\sqrt{5}}{10}}
-1 \right) \quad .
\end{eqnarray}
Within the linearized approximation,
the distribution $\omega(c_2,c_3,t)$ has its maximum located in 
$\hat c_2 (t), \hat {c}_3 (t)$ with a height $\hat \omega (t)$, and
is equal to minus infinity elsewhere. The coordinates of the maximum 
as functions of the depth $t$ defines the tree trajectory, {\em i.e.} 
the evolution, driven by the action of DPLL, of the dominant branches 
in the phase diagram of Figure~\ref{diag}.  We obtain straightforwardly
this trajectory from equations (\ref{c3c}) and the transformation 
rules (\ref{rules}). 

\subsubsection{Interpretation of the tree trajectories and results for 
the complexity.}

In Figure~\ref{diag}, the tree trajectories corresponding to solving
3-SAT instances with ratios $\alpha_0=4.3,7$ and $10$ are shown. The
trajectories start on the right vertical axis $p=1$ and head to the
left until they hit the halt line $\alpha\simeq 1.259/(1-p)$ 
(dot-dashed curve) at some time $t_h >0$, which
depends on $\alpha _0$. On the halt line, a
delocalization transition for the largest eigenvector takes place (for
parameters $y_2=y_3=0$, see Appendix~\ref{diagosing} and
Figure~\ref{m2critcourbe}) and causes an avalanche of unitary clauses
with the emergence of contradictions, preventing branches from further
growing.

The delocalization transition taking place on the halt line means that
the stationary probability $\mu _t (0)$ of having no unit-clause
\begin{equation}
\mu _t (0) = \frac { \tilde v _0 (0)}{\sum _{C_1=0}^\infty
\tilde v _0 (C_1) } = \frac{ \tilde V _0(0)} { \tilde V _0(1)} 
\qquad , \label{mu0det}
\end{equation}
vanishes at $t=t_h$. From equations 
(\ref{eigenva1},\ref{eiggf2},\ref{mu0det},\ref{eiggf3},\ref{eiggf}),
the largest eigenvalue for dominant branches, $\Lambda (0,0)$ reaches its
lowest value, one, on the  halt line.  As expected,
the emergence of contradictions on dominant branches coincides with
the halt of the tree growth, see equation (\ref{parf}).

The logarithm $\hat \omega (t)$ of the number of dominant branches 
increases along the tree trajectory, from zero at $t=0$ up to 
some value $\hat \omega (t_h) > 0$ on the halt line.
This final value, divided by $\ln 2$, 
is our analytical prediction (within the linear
approximation of Section~\ref{yhn}) for the complexity $\omega$. 
We describe in Appendix~\ref{quadapp}, a refined, quadratic expression
for the Legendre transform $\varphi$~(\ref{vp}) that provides another
estimate of $\omega$. 

The theoretical values of $\omega$, within linear and quadratic approximations,
are shown in Table~\ref{tabunsat} for $\alpha_0=20,15,10,7,4.3$,
and  compare very well with numerical
results. Our calculation, which is fact an annealed estimate of $\omega$ 
(Section~\ref{foc}), is very accurate. The decorrelation 
approximation (Section~\ref{seces}) becomes 
more and more precise with larger
and larger ratios $\alpha_0$. Indeed, the probability that the same variable
appears twice in the search tree decreases for smaller trees.  
For large values of $\alpha _0$, we obtain
\begin{equation} 
\omega (\alpha _0) \sim
\frac{(3+\sqrt{5})\left[ \ln \left( \frac{1+\sqrt{5}}{2} \right)
\right]^2}{6\, \ln 2\;\alpha _0}
\sim \frac{0.292}{\alpha _0}
\end{equation}
The $1/\alpha _0$ 
scaling of $\omega$ has been previously proven by
Beam, Karp, and Pitassi \cite{Bea}, independently of the particular 
heuristics used.  Showing that there
is no solution for random 3-SAT instances with large ratios $\alpha_0$
is relatively easy, since assumptions on few variables generate a large
number of logical consequences, and contradictions emerge quickly, 
see Figure~\ref{tempi}. This result can be inferred from 
Figure~\ref{diag}. As $\alpha _0$ increases, the
distance between the vertical 3-SAT axis and the halt line
decreases; consequently, the trajectory become
shorter, and so does the size of the search tree.

\subsubsection{Length of the dominant branches.} 

To end with, we calculate the length of the dominant branches.
The probability that a splitting occurs at time $t$ is $\mu _t (0)$
defined in (\ref{mu0det}). Let us define $n ( L , T)$ as the number
of branches having $L$ nodes at depth $T$ in the tree.
The evolution equation for $n$ is
$n(L,T+1)=(1-\mu _t(0) )\,n(L,T)+2\,\mu_t(0)\; n(L-1,T)$.
The average branch length at time $T$,
$\langle L (T) \rangle = {\sum_{L=1} ^\infty \, L\, n(L,T)}/{\sum_ {L=0}
^\infty n(L,T)}$,
obeys the simple evolution relation
$\langle L (T+1) \rangle - \langle L (T) \rangle= 
2\,\mu_t (0)/(1+\mu _t (0))$.
Therefore, the average number of nodes (divided by $N$), $l=
\langle L (t\,N) \rangle/N$, present
along dominant branches once the tree is complete, is equal to
\begin{equation}
\langle \ell \rangle =\int_0^{t_{h}}\,dt\;\frac{2\,\mu _t (0)}{1+\mu _t(0)}
\end{equation}
where $t_h$ is the halt time. For large ratios $\alpha _0$, the 
average length of dominant branches scales as
\begin{equation}
\langle \ell \rangle (\alpha _0) \sim
\frac{(3+\sqrt{5})\ln \left( \frac{1+\sqrt{5}}{2} \right)
 +1 -\sqrt{5} }{3\;\alpha _0}
\sim \frac{0.428}{\alpha _0} \qquad .
\end{equation}
The good agreement between this prediction and the numerical results
can be checked on the insets of Figure~\ref{histo} for different
values of $\alpha_0$.

\section{Mixed trajectories and the intermediate exponential regime
(upper sat phase).} 
\label{secmixed}

In this section, we show how the complexity of solving 3-SAT
instances with ratios in the intermediate range
$\alpha _L < \alpha _0 < \alpha _C$  can be understood by combining
the previous results on branch and tree trajectories. 

\subsection{Branch trajectories and the critical line of 2+p-SAT.}
\label{tele}

In the upper sat phase, the single branch trajectory intersects the critical
line $\alpha _c(p)$ in some point $G$, whose coordinates depend on
the initial ratio $\alpha_0$. The point $G$ corresponding
to $\alpha _0=3.5$ is shown in Figure~\ref{diag}.

For a finite size $N$, the critical 2+p-SAT region (also called critical
window) around $G$ has a non zero width $W$ in terms of the numbers of
clauses and variables, much smaller than $N$, since the transition is
sharp\cite{Gut}. Let us call G$_-$ and G$_+$ the lower and upper
borders of this windows along the first branch run by DPLL, see bold
line on Figure~\ref{treeinter}. We also denote by $N_-$ and $N_+$ the 
average numbers of variables not assigned by DPLL at points G$_-$ and G$_+$
respectively: $N_+ < N_- < N$. G$_-$ carries an unsat 2+p-SAT instance.
A refutation subtree must be built by DPLL before backtracking above $G_-$. 
The corresponding (sub)tree trajectory, starting
from $G$ and penetrating the unsat phase up to the arrest line, is shown
on Figure~\ref{diag}.

The size $2^{N_- \omega _-}$ of the subtree obviously provides 
a lower bound to the total 
complexity. Now, once the subtree has been entirely
explored, DPLL backtracks to some node lying above $G_-$ in the tree
(Figures~\ref{tree} and \ref{treeinter}). The highest backtracking
node, say G$_0$, is necessarily the deepest one (when starting from
above) along the first DPLL branch that carries a satisfiable
2+p-instance, and lies below G$_+$.  Therefore, a solution must
necessarily be found by DPLL below G$_+$. 
The corresponding branch (rightmost path in Figure~\ref{tree}C) 
is highly non typical and does not contribute
to the complexity, since almost all branches in the search tree
are described by the tree trajectory issued from G (Figure~\ref{diag}).
The total size of the search
tree is thus bounded from above by $2^W \times 2^{N_- \omega _-}$ and, 
to exponentially dominant order, equivalent to the size of the subtree 
below G$_-$.

\subsection{Analytical calculation of the size of the refutation subtree.}

The coordinates $p_G, \alpha _G = \alpha _C (p_G)$ of the crossing point 
$G$ depend on the initial 3-SAT ratio $\alpha_0$ and may be computed 
from the knowledge of the 2+p-SAT critical line $\alpha _C (p)$ and 
the branch trajectory equations (\ref{eqsup}). For $\alpha _0=3.5$,
we obtain $p_G=0.78$ and $\alpha_G=3.02$. Point $G$ is reached by the
branch trajectory once a fraction $t_G \simeq 0.19$ of variables
have been assigned by DPLL.

Once G is known, we consider the unsatisfiable 2+p-SAT instances
with parameters $p_G, \alpha _G$ as a starting point for DPLL. The
calculation exposed in Section~\ref{sectree} can be used with initial
conditions $p_G,\alpha _G$.  We show in Table~\ref{tabmixed} 
the results of the analytical calculation of $\omega _G$,
within linear and quadratic approximations for
a starting ratio $\alpha _0 =3.5$. Note that the discrepancy between
both predictions is larger than for higher values of $\alpha _0$.

The logarithm $\omega$ of the total complexity is defined through the
identity $2^{N\,\omega}=2^{N_G\;\omega _G}$, or equivalently,
\begin{equation}
\label{erreur}
\omega =  \omega _G \; (1-t_G) \qquad .
\end{equation}
The resulting value for $\alpha _0=3.5$ is shown in Table~\ref{tabmixed}.

\subsection{Comparison with numerics for $\alpha _0=3.5$.}

We have checked numerically the scenario of Section~\ref{tele} in two 
ways. 

First, we have computed during the action of DPLL, the coordinates in
the $p,\alpha$ plane of the highest backtracking point in the search
tree. The agreement with the coordinates of G computed in the
previous paragraph is very good (Section~\ref{secnum}). However,
the numerical data show large fluctuation and the experimental fits
are not very accurate, leading to uncertainties on $p_G$ and $\alpha
_G$ of the order of $0.01$ and $0.02$ respectively. In addition, note
that the analytical values of the coordinates of $G$ are not
exact since the critical line $\alpha _C (p)$ is not
rigorously known (Section~\ref{deuxpp}).

Secondly, we compare in Table~\ref{tabmixed} the
experimental measures and theoretical predictions of the complexity
starting from $G$. The agreement between all values is quite good
and lead to a complexity about $\omega _G = 0.042 \pm 0.002$. Numerics
indicate that the annealed value of the complexity is equal (or
slightly larger) than the typical value. Therefore the annealed calculation
developed in Section~\ref{sectree} agrees well the data obtained for 
2+p-SAT instances. Once $\omega _G$ and $t_G$ are known, eqn.(\ref{erreur})
gives access to the theoretical value of $\omega$. 

The agreement between theory and experiment is very satisfactory 
(Table~\ref{tabmixed}). Nevertheless, let us stress the existence of
some uncertainty regarding the values of the highest backtracking point
coordinates $p_G, \alpha _G$. Numerical simulations on 2+p-SAT
instances and analytical checks show that $\omega$ depends strongly on
the initial fraction
$p_0$.  Variations of the initial parameter $p_G$ around 0.78 by
$\Delta p_0 = 0.01$ change the final result for the complexity by
$\Delta \omega = 0.003 - 0.004$, twice as large as the statistical
uncertainty at fixed $p_0=p_G=0.78$. Improving the accuracy of the
data would require a precise determination of the coordinates of G.

We show in Figure~\ref{diag} the trajectory of the atypical, rightmost branch 
(ending with a solution) in the tree, 
obtained from simulations for $N=300$. It comes as
no surprise that this trajectory, which carries a satisfiable and
highly biased 2+p-SAT instance, may enter the unsat region defined
for the unbiased 2+p-SAT distribution. The trajectory eventually reaches
the $\alpha =0$ axis when all clauses are eliminated. Notice that the end
point is not S, but the lower left corner of the phase diagram.

As a conclusion, our work shows that, in the $\alpha_L < \alpha < \alpha
_C$ range, the complexity of solving 3-SAT is related to the existence
of a critical point of 2+p-SAT. The right part of the 2+p-SAT critical line,
comprised between T and the threshold point of 3-SAT, can be determined
experimentally as the locus of the highest backtracking points in 3-SAT solving
search trees, when the starting ratio $\alpha_0$ spans the interval $\alpha _L
\le \alpha_0 \le \alpha _C$.

\section{Complexity of 2+p-SAT solving and relationship with static.}
\label{secpattern}

\subsection{Complexity diagram.}

We have analyzed in the previous sections the computational complexity of
3-SAT solving. The analysis may extended to any 2+p-SAT instance, with
the results shown in Figure~\ref{pattern}.

In addition to the three regimes unveiled in the study of 
3-SAT\footnote{Though differential equations (\ref{diff}) depend
on $t$ when written in terms of $c_2, c_3$, they are Markovian if
rewritten in the variables $p,\alpha$. Therefore, the locus of the 
2+p-SAT instances points, $p_0,\alpha _0$, giving rise to trajectories
touching the threshold line in T,
simply coincides with the 3-SAT trajectory
starting from $\alpha_0=\alpha _L$.}, a new
complexity region appears, on the left side of the line $\alpha =1/(1-p)$,
referred to as ``weak delocalization'' line. To the right (respectively
left) of the weak delocalization line, the second largest eigenvector 
of the branching matrix $K$ is localized (resp. delocalized), see 
Appendix~\ref{diagosing}. When solving a 3-SAT instance, or more generally
a 2+p-SAT  instance with parameters $p_0, \alpha _0 \le 1/(1-p_0)$, 
the size of the search tree is exponential in $N$ when the weak delocalization
line is crossed by the tree trajectory (Figure~\ref{diag}). 
Thus, the contribution to the
average flow of unit-clauses coming from the second largest
eigenvector is exponentially damped by the largest eigenvector contribution, and
no contradiction arises until the halt, strong delocalization line is hit.

If one now desires to solve a 2+p-SAT instance whose representative point 
$p_0,\alpha_0$ lies to the left of the weak delocalization curve, the 
delocalization of the second largest eigenvector stops immediately
the growth of the tree, before the distribution of 1-clauses could reach
equilibrium, see discussion of Section~\ref{sec57}. 
Therefore, in the range of parameters $p_0, \alpha_0 \ge 1/(1-p_0)$, 
proving unsatisfiability does not require an exponentially
large computational effort. 

\subsection{Polynomial/exponential crossover and the tricritical point.}

The inset of Figure~\ref{pattern} show a schematic blow up of the 
neighborhood of T and T$_S$, where all complexity regions meet. 
From the above discussion, the complexity of solving critically constrained
2+p-SAT instances is polynomial up to $p_S$, and exponential above,
in agreement with previous claims based on numerical investigations
\cite{Sta}. In the range $2/5 < p_0 < p_S$, computational complexity
exhibits a somewhat unusual behavior as a function of $\alpha$. 
The peak of hardness is indeed not located at criticality (where
the scaling of complexity is only polynomial), but
slightly below threshold, where complexity is exponential. Unfortunately,
the narrowness of the region shown in the Inset of Figure~\ref{pattern}
seems to rule out the possibility of checking this statement through
experiments.

To end with, let us stress that T, conversely to T$_S$, depends {\em a priori}
on the splitting  heuristic. Nevertheless, the location of
T seems to be more insensitive to the choice of the heuristics than branch 
trajectories. For instance, the UC and GUC heuristics both lead to the
same tangential hit point T, while the starting ratios of the corresponding
trajectories, $\alpha _L =8/3$ and $\alpha _L \simeq 3.003$, differ.
Understanding this relative robustness, and the surprising closeness
of T and T$_S$, would be interesting.

\section{Conclusion and perspectives.}
\label{secconc}

\begin{center}
\begin{figure}
\includegraphics[width=300pt,angle=-90]{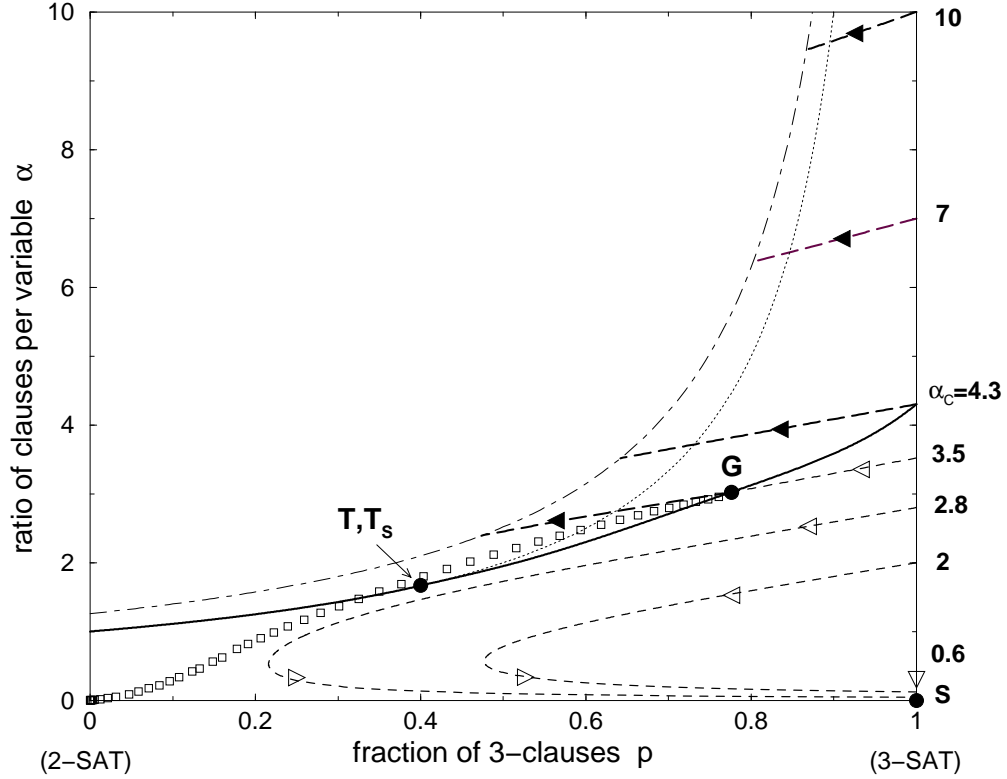}
\vskip .5cm
\caption{Phase diagram of the 2+p-SAT model and trajectories generated
by DPLL.
The threshold line $\alpha_C (p)$ (bold full line) separates
sat (lower part of the plane) from unsat (upper part)
phases. Extremities lie on the vertical 2-SAT (left) and 3-SAT (right)
axis at coordinates ($p=0,\alpha _C=1$) and ($p=1,\alpha _C\simeq
4.3$) respectively. The threshold line coincides with the $\alpha
=1/(1-p)$ curve (dotted line) when $p \le  p_S \simeq 0.41$, that is,
up to the tricritical point T$_S$. 
Departure points for DPLL trajectories are located on the 3-SAT
vertical axis and the corresponding values of $\alpha _0$ are explicitly
given. Dashed curves represent tree trajectories in the unsat region
(thick lines, black arrows) and branch trajectories in the
sat phase (thin lines, empty arrows). Arrows indicate the direction of
motion along trajectories parametrized by the fraction $t$ of
variables set by DPLL.  For small ratios, e.g.  $\alpha _0 =2 < \alpha
_L\simeq 3.003$, 
trajectories remain confined in the sat phase. At $\alpha_L$, the
single branch trajectory hits tangentially the threshold line in T of
coordinates $(2/5,5/3)$, very close to T$_S$.
In the intermediate range $\alpha _L < \alpha
_0 < \alpha_C$, the branch trajectory intersects the threshold line at
some point G that depends on $\alpha$. 
A dense tree then grows in the unsat phase, as happens when 3-SAT
departure ratios are above threshold $\alpha > \alpha _C \simeq
4.3$. The tree trajectory halts on the dot-dashed curve $\alpha \simeq
1.259/(1-p)$ where the tree growth process stops.  At this
point, DPLL has reached back the highest backtracking node in the
search tree, that is, the first node when $\alpha _0> \alpha _C $, or
node G for $\alpha _L < \alpha _0< \alpha_C$.  In the latter case, a
solution can be reached from a new descending branch 
(rightmost path in Figure~\ref{tree}C)
while, in the former case, unsatisfiability is proven. Small squares
show the trajectory corresponding to this successful branch for $\alpha=3.5$,
as obtained from simulations for $N=300$. The trajectory coincides perfectly
with the theoretical branch trajectory up to point G (not shown), and then
reaches the $\alpha =0$ axis when a solution is found. 
}
\label{diag}
\end{figure}
\end{center}

\begin{center}
\begin{figure}
\includegraphics[width=170pt,angle=-90]{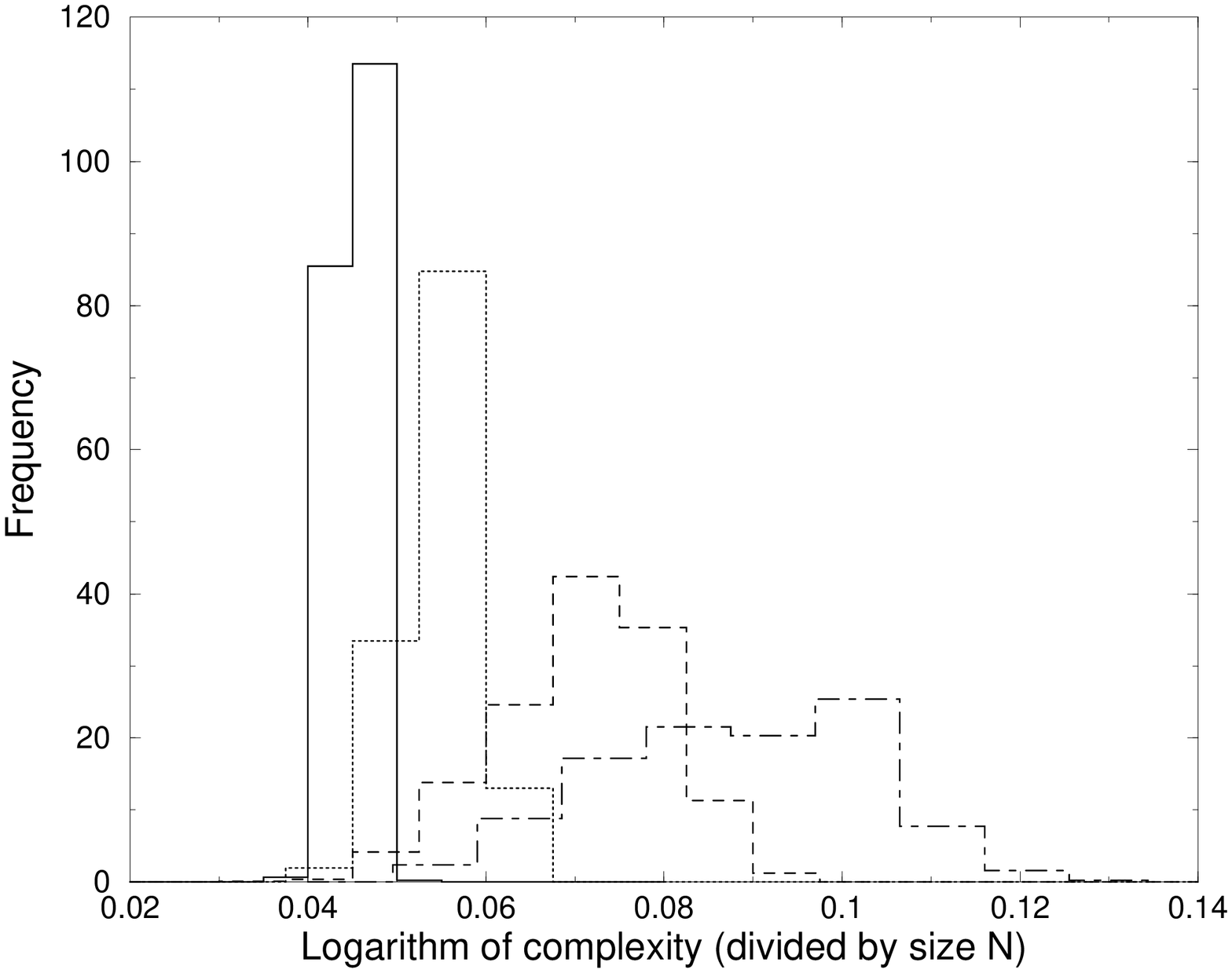}
\includegraphics[width=170pt,angle=-90]{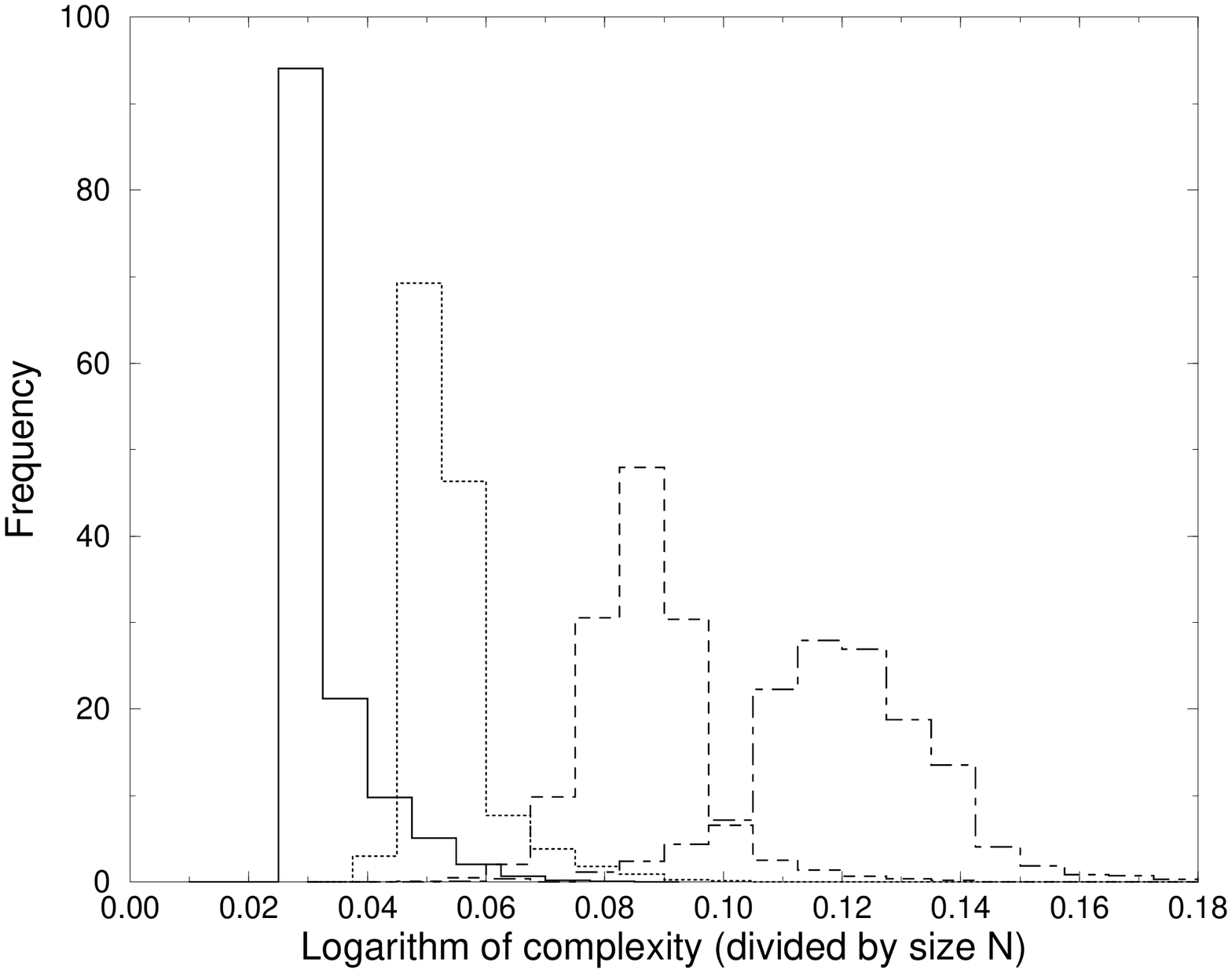}
\vskip 1cm
\caption{Distribution of the logarithms (in base 2, and divided by
$N$) of the complexity for four different sizes $N=$ 30
(dot-dashed), 50 (dashed), 100 (dotted) and 200 (full line). The
histograms are normalized to unity and represent 50,000 instances. 
The distribution gets more and more concentrated as the size grows.
{\bf A.} Ratio $\alpha  =10$. Curves are roughly symmetrical around their
mean with small tails on their flanks; 
large fluctuations from sample to sample are absent.
{\bf B.} Ratio $\alpha  =3.1$. Curves have large tails on the right,
reflecting the presence of rare, very hard samples.}
\label{histoseuil}
\end{figure}
\end{center}

\begin{center}
\begin{figure}
\includegraphics[width=250pt,height=330pt,angle=-90]{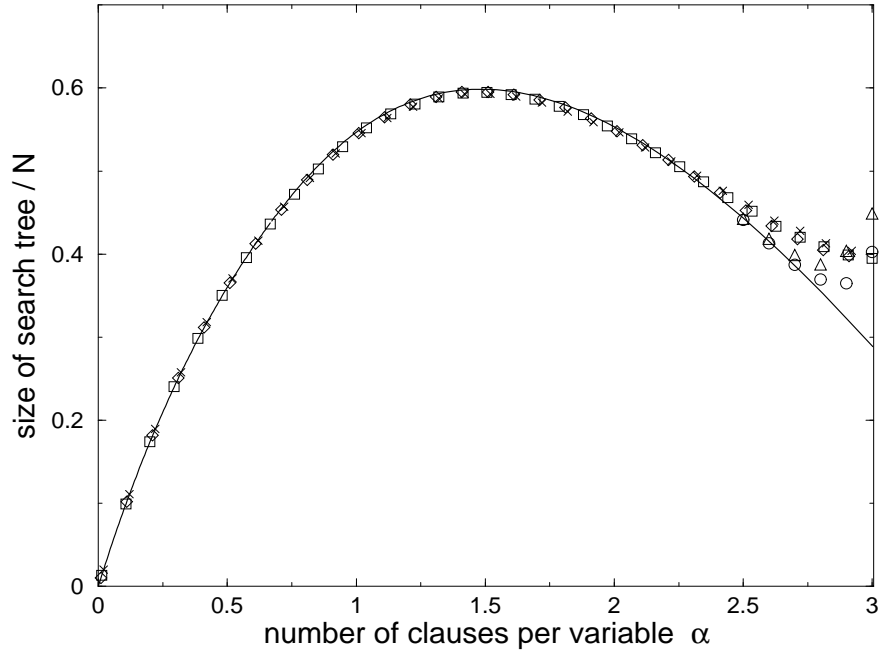}
\vskip .5cm
\caption{Complexity of solving in the sat region for $\alpha < \alpha
_L \simeq 3.003$, divided by the size $N$ of the instances. Numerical
data are for sizes $N=$50 (cross), 75 (square), 100 (diamond), 500
(triangle) and 1000 (circle). For the two biggest sizes, simulations
have been carried out for ratios larger than 2.5 only. Data for
different $N$ collapse onto the same curve, proving that complexity 
scales linearly with $N$. The bold continuous curve is the analytical
prediction $\gamma (\alpha)$ 
from Section~\ref{length}. Note the perfect agreement with
numerics except at large ratios where finite size effects are
important, due to the cross-over to the exponential regime above
$\alpha _L \simeq 3.003$.}
\label{constante}
\end{figure}
\end{center}

\begin{center}
\begin{figure}
\includegraphics[height=350pt,width=205pt,angle=-90]{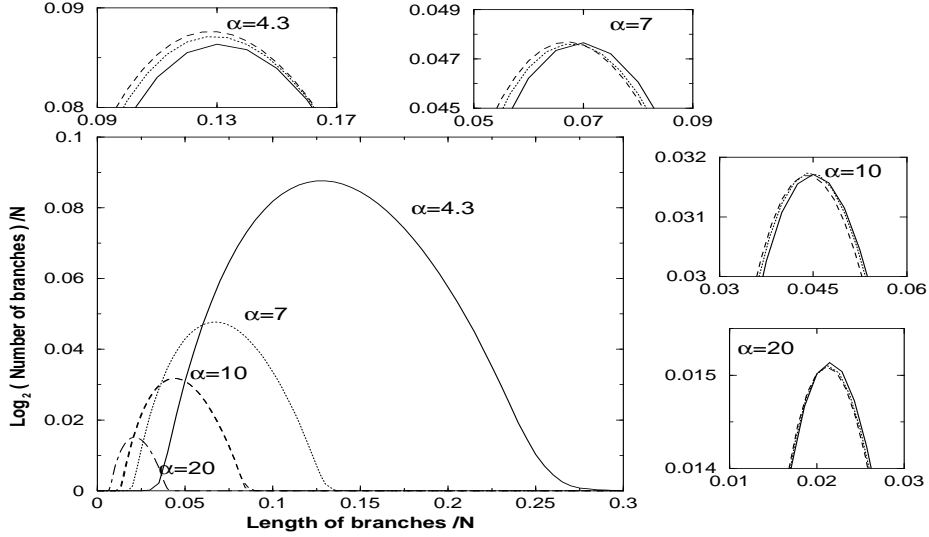}
\vskip 1cm
\caption{Logarithm of the number of branches in unsat search trees as
a function of the branch length. Main figure: the size of the search 
tree is a decreasing function of $\alpha$ at fixed $N$. Histograms are
presented here for ratios equal to $\alpha=4.3$ (solid line, $N=200$), 
$\alpha=7$ (dotted line, $N=400$), $\alpha=10$ (dashed line, $N=600$)
and $\alpha=20$ (dotted-dashed line, $N=900$) and have been averaged
over hundreds of samples.
Inset: the heights of the tops of the histograms show a very weak 
dependence upon $N$. 
Numerical extrapolations of $\omega_N$ to $N\to \infty$ and statistical errors 
are reported Table~\ref{tabunsat}. Sat instances (which may be present
for small sizes at $\alpha =4.3$) have not be considered in the averaging
procedure. For each inset, we give below the sizes $N$ followed
by the number of instances in parenthesis used for averaging.
Ratio $\alpha=4.3$: solid line: 100 (5000), dotted line: 150 (500), 
dashed line: 200 (400); 
$\alpha=7$: solid line: 200 (10000), dotted
line: 300 (1000), dashed line: 400 (200).
$\alpha=10$: solid line: 400 (500), dotted line:
500 (400), dashed line: 600 (100);
$\alpha=20$: solid line: 700 (600), dotted
line: 800 (1000), dashed line: 900 (1000).  }
\label{histo}
\end{figure}
\end{center}

\begin{center}
\begin{figure}
\includegraphics[width=160pt,angle=-90]{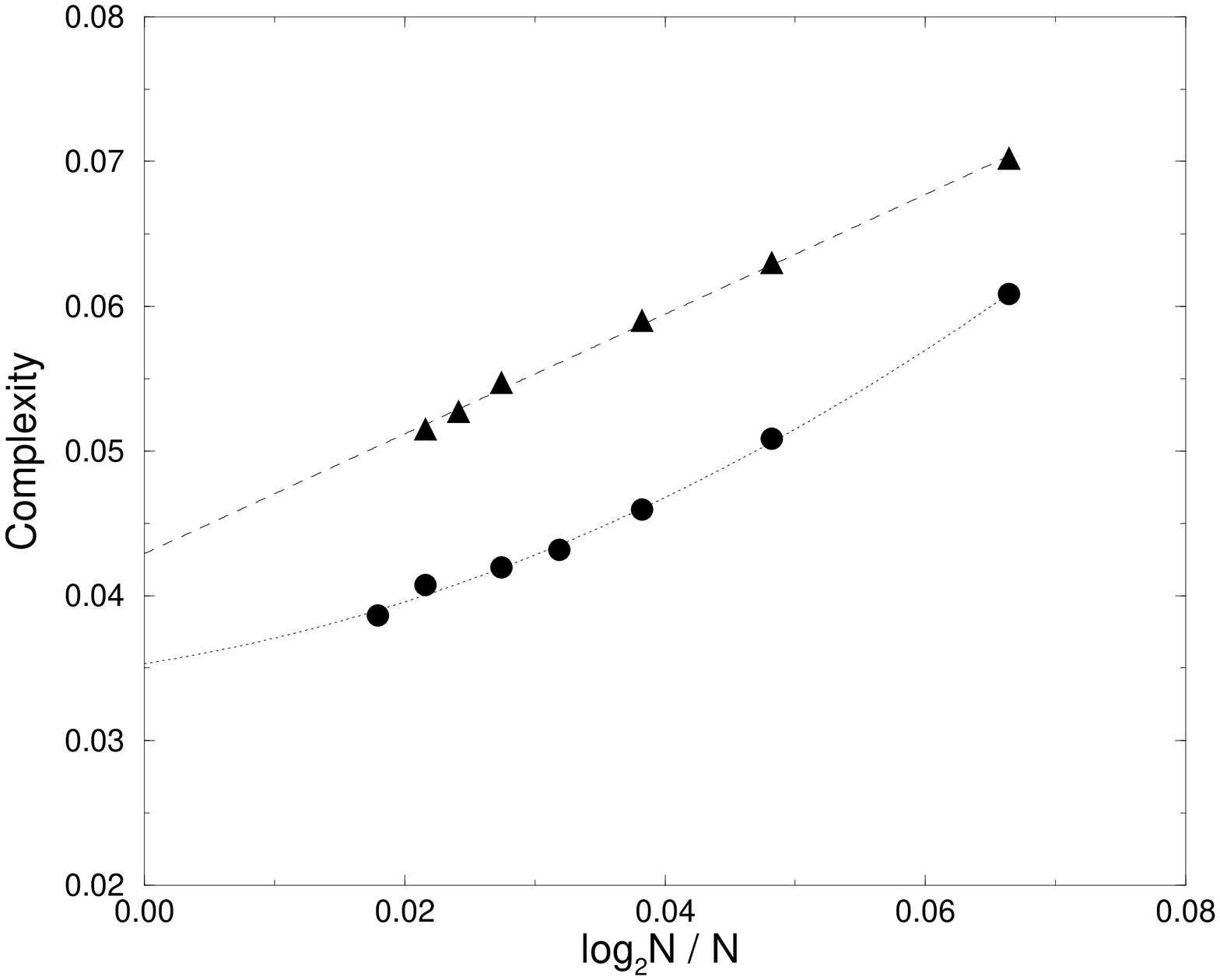}
\includegraphics[width=160pt,angle=-90]{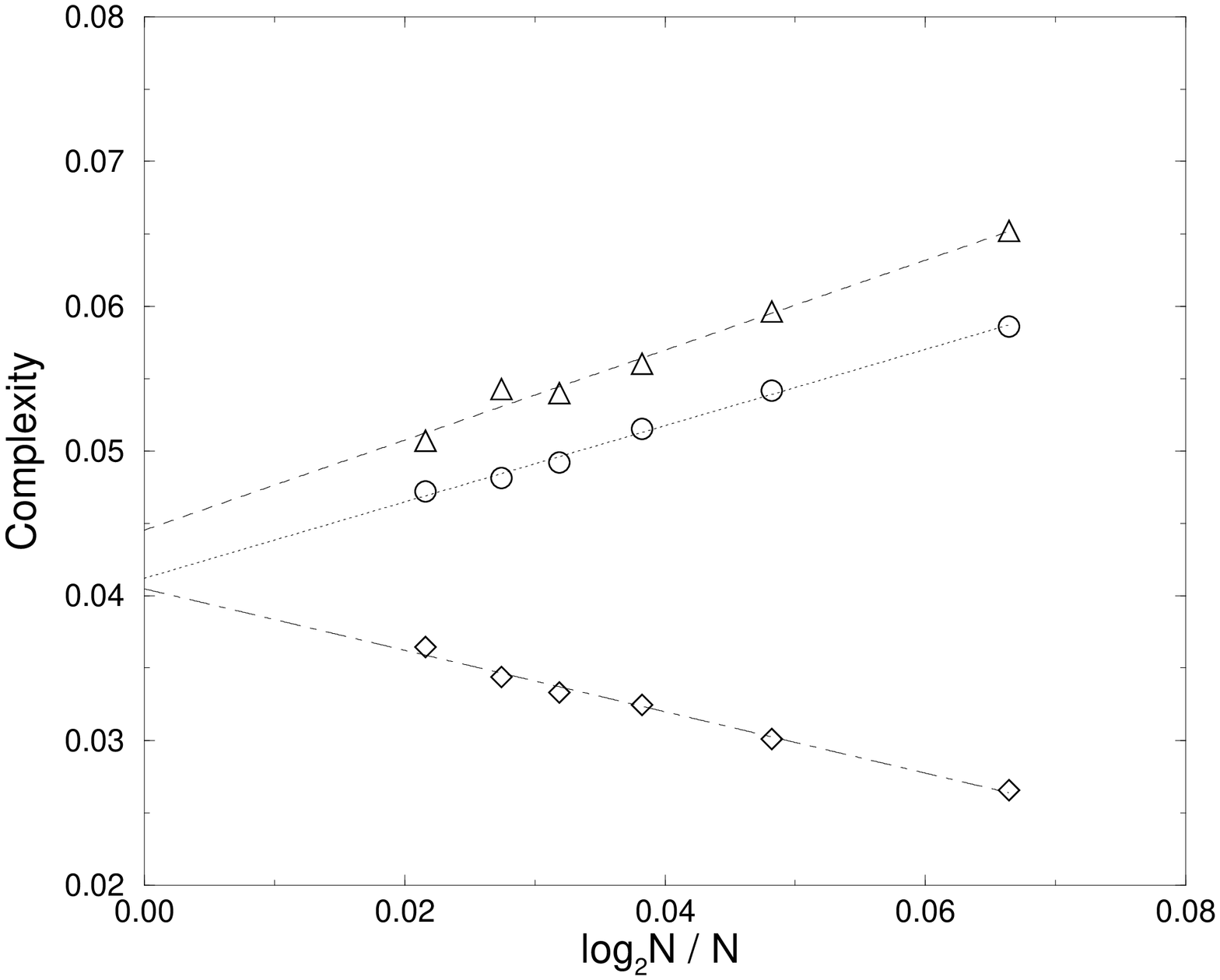}
\vskip 1cm
\caption{Solving complexity in the upper sat phase for sizes $N$ ranging 
from $N=100$ to $N=400$. The size of the symbols accounts for the largest
statistical error bar. {\bf A.} 3-SAT problem with ratio $\alpha =3.5$:
typical (average of the logarithm, full circles) and
annealed (logarithm of the average, full triangles) size of the
search tree. Dotted lines are quadratic and linear fit of the typical and
annealed complexities, giving $\omega _{3} ^{typ} = 0.035 \pm 0.03$
and $\omega _3 ^{ann} = 0.043 \pm 0.02$ in the infinite size limit. {\bf B.} 
Related 2+p-SAT problem with parameters $p_G = 0.78, 
\alpha _G=3.02$. The typical complexity is measured from the size
of the search tree (circles) and the top of the branch length distribution 
(triangle), with the same large $N$ extrapolation: $\omega _{2+p} ^{typ} =
0.041 \pm 0.02$. This value is slightly smaller than
the annealed complexity $\omega _{2+p} ^{ann} = 0.044 \pm 0.02$. 
All fits are linear.}
\label{comp3.5}
\end{figure}
\end{center}

\begin{center}
\begin{figure}
\includegraphics[width=160pt,angle=-90]{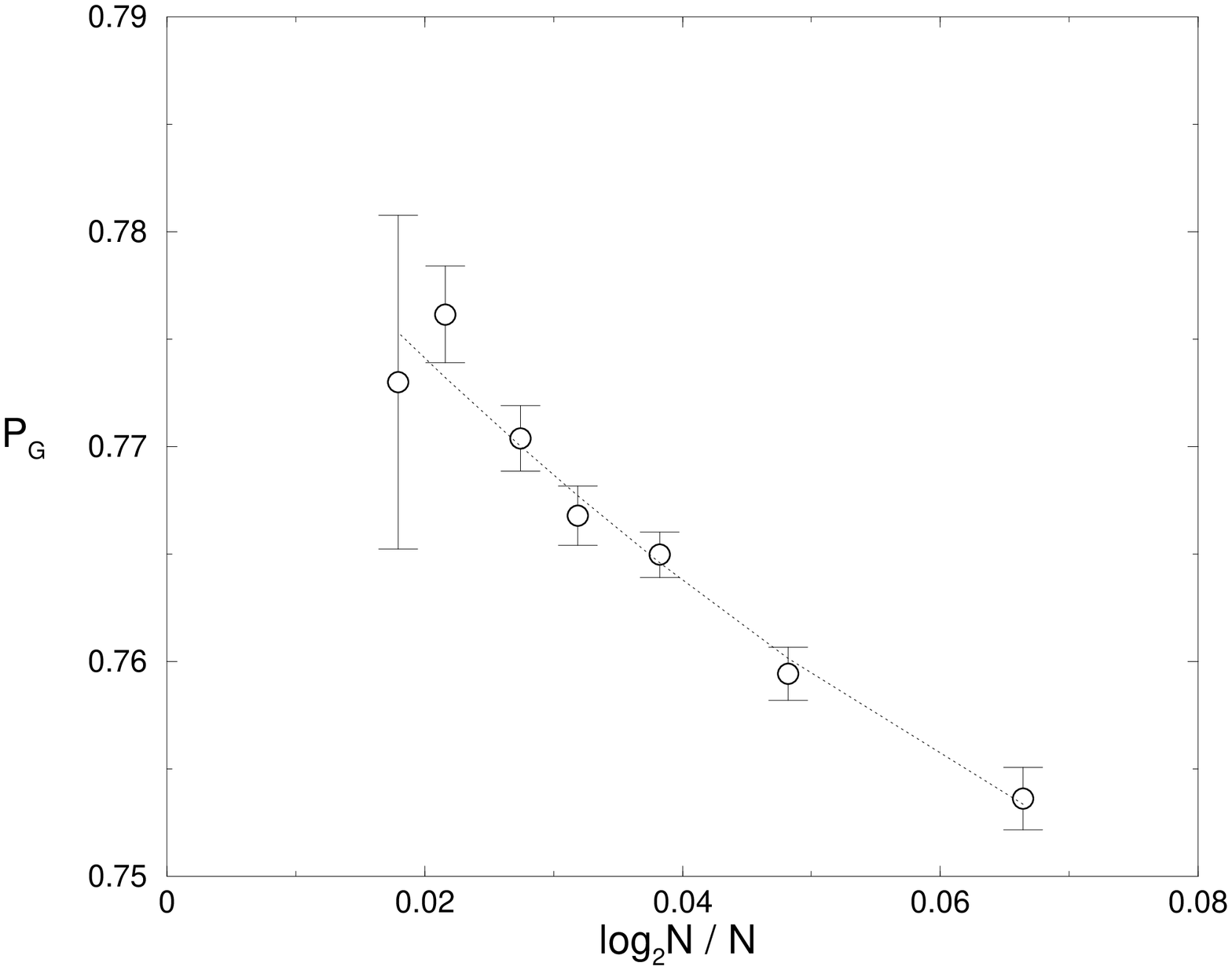}
\includegraphics[width=160pt,angle=-90]{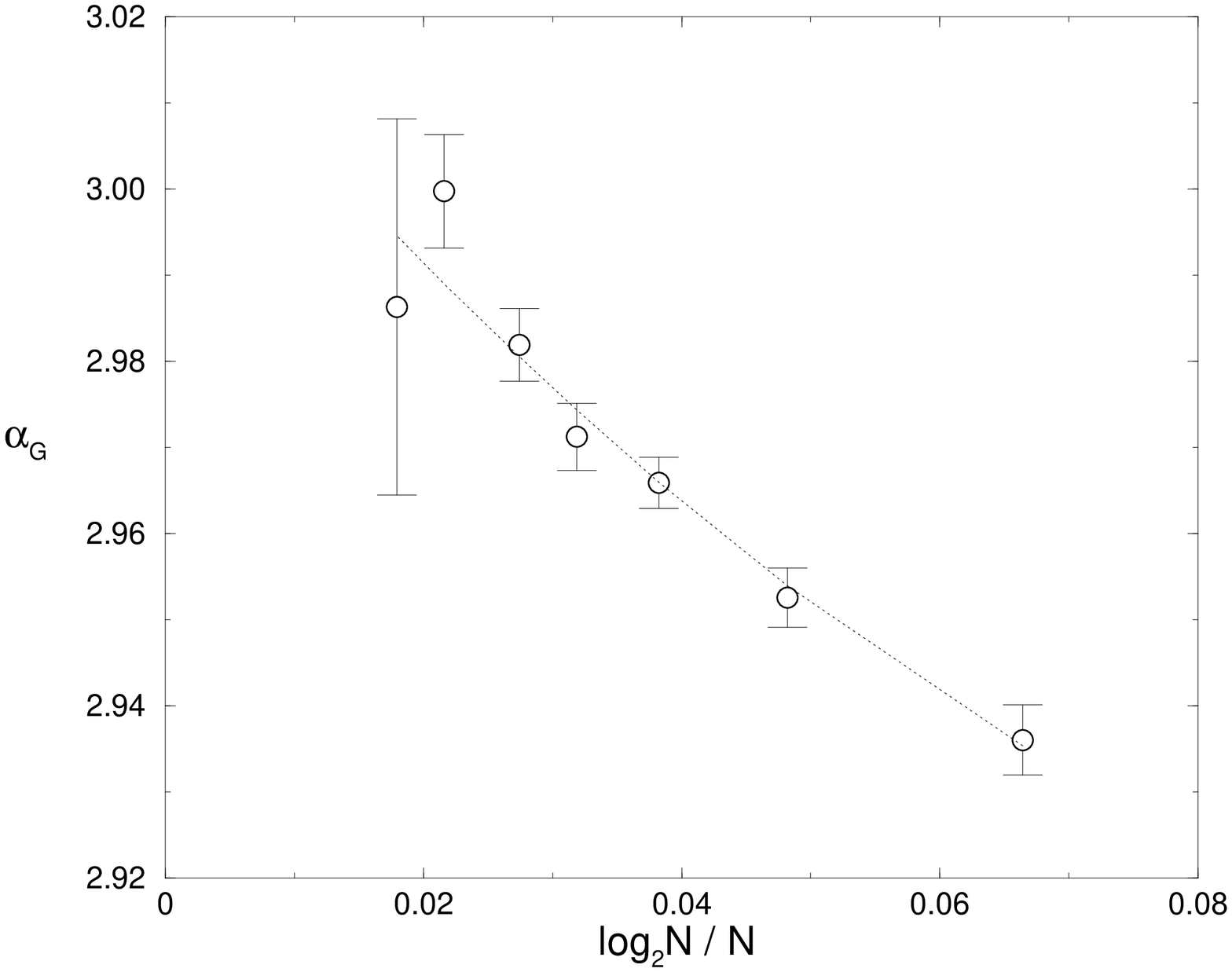}
\vskip 1cm
\caption{Coordinates of the highest backtracking point G in the search
tree with a starting ratio $\alpha =3.5$ (the upper sat phase) 
for different sizes $N=100, \ldots , 500$ and averaged over 10,000 (small
sizes) to $128$ ($N=500$) instances. The fits shown are quadratic functions
of the plotting coordinate $\log _2 N/N$, and give the extrapolated
location of $G$ in the large size limit: $p_G = 0.78 \pm 0.01$, $\alpha _G
=\ 3.02 \pm 0.02$. Note the uncertainty on these values due to the
few number of instances available at large instance sizes. }
\label{pgalphag}
\end{figure}
\end{center}

\begin{center}
\begin{figure}
\includegraphics[width=250pt,angle=-0]{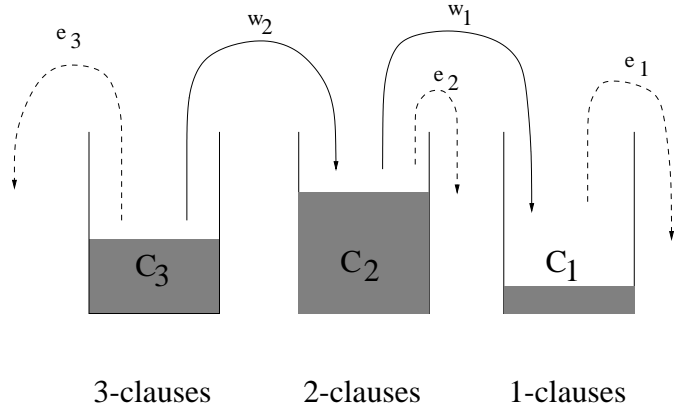}
\vskip .5cm
\caption{Schematic view of the dynamics of clauses. Clauses are sorted
into three recipients according to their lengths, {\em i.e.} the
number of variables they include. Each time a variable is assigned by
DPLL, clauses are modified, resulting in a dynamics of the recipients
populations (lines with arrows). Dashed lines indicate the
elimination of (satisfied) clauses of lengths 1, 2 or 3. Bold lines
represent the reduction  of 3-clauses into 2-clauses, or 2-clauses into
1-clauses. The flows of clauses 
are denoted by $e_1, e_2, e_3$ and $w_2, w_1$ respectively.
A solution is found when all recipients are empty.  The level of the
rightmost recipient coincides with the number of unitary clauses. If
this level is low ({\em i.e.} $O(1)$), the probability that two
contradictory clauses $x$ and $\bar x$ are present in the recipient is
vanishingly small.  When the level is high ({\em i.e.} $O(\sqrt N)$),
contradictions will occur with a large probability. }
\label{recip}
\end{figure}
\end{center}

\begin{center}
\begin{figure}
\includegraphics[width=170pt,angle=0]{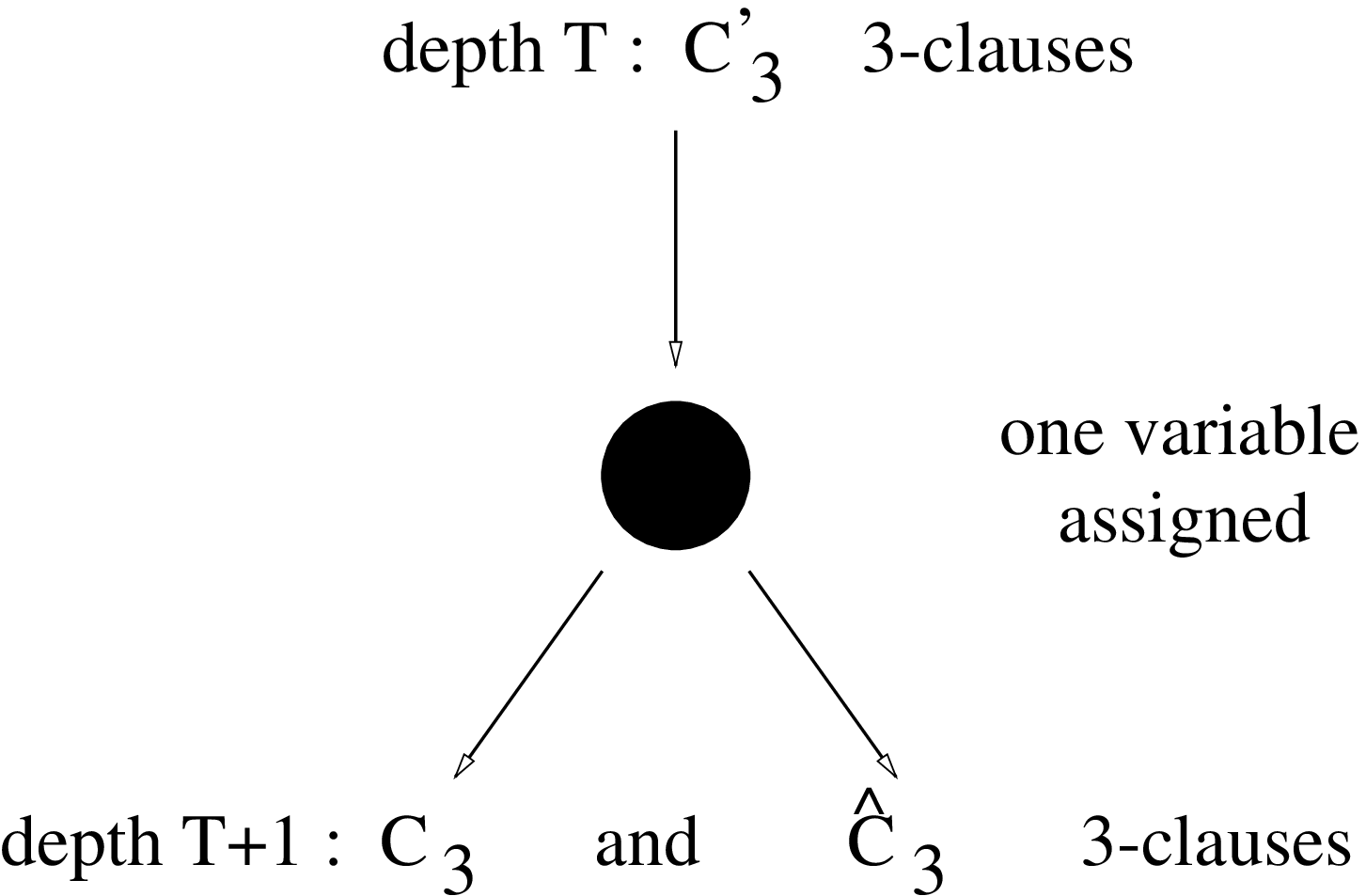}
\vskip .5cm
\caption{Schematic representation of the search tree building up in the
toy dynamical process. A branch carrying a formula with $C'_3$ 3-clauses
splits once a variable is assigned, and gives birth to two branches 
having $C_3$ and $\hat C_3$ 3-clauses
respectively. The variable is assigned randomly, independently of the 
3-clauses. Clauses of lengths one or two are not considered.}
\label{algotoy}
\end{figure}
\end{center}

\begin{center}
\begin{figure}
\includegraphics[width=170pt,angle=-90]{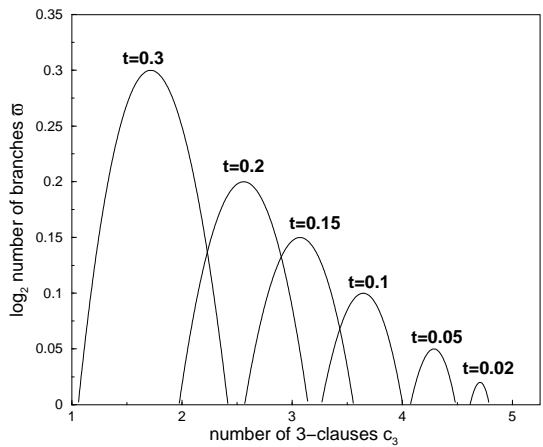}
\vskip .3cm
\caption{Logarithm $\omega$ of the number of branches (base 2,
and divided by $N$) as a function
of the number $c_3$ of 3-clauses at different times $t$ for the simplified
growth process of Section~\ref{seces}. At time $t=0$, the search tree is 
empty and the ratio of clauses per variable equals $\alpha _0 =5$.
Later on, the tree grows and a whole distribution of branches 
appears. Dominant branches correspond to the top of the distributions
of coordinates $\hat c_3 (t), \hat \omega  (t)$. Branches
become exponentially more numerous with time, while they carry
less and less 3-clauses.}
\label{entroy}
\end{figure}
\end{center}

\begin{center}
\begin{figure}
\includegraphics[width=400pt,height=250pt,angle=0]{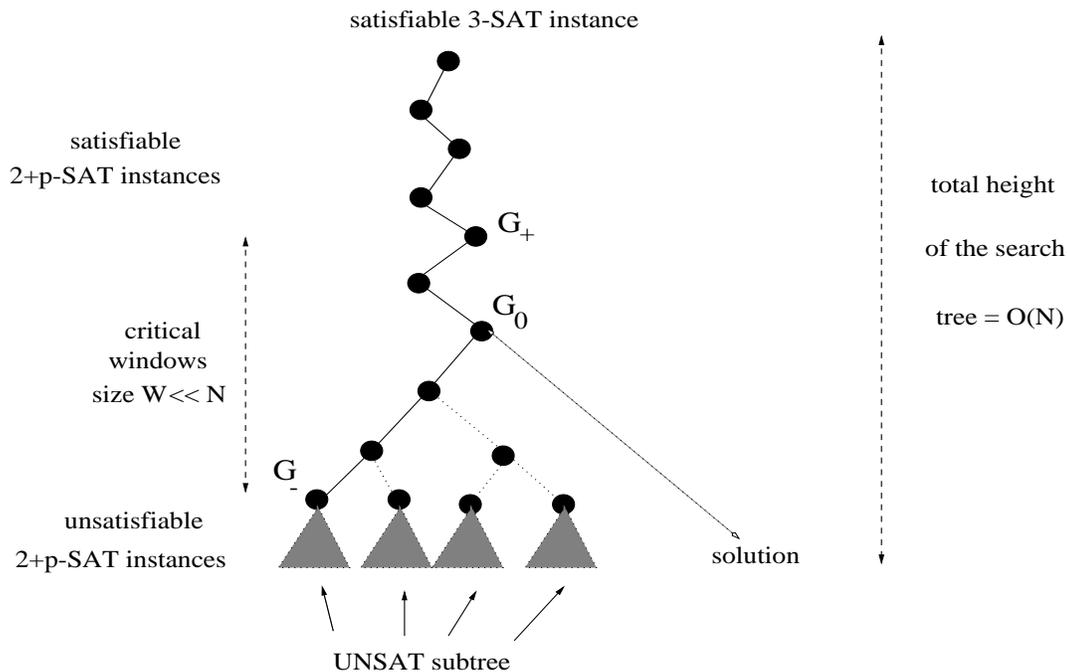}
\vskip .3cm
\caption{Detailed structure of the search tree in the 
upper sat phase ($\alpha _L < \alpha < \alpha _C$). DPLL starts with a 
satisfiable 3-SAT instance and transforms it into a sequence of 
2+p-SAT instances. The bold, leftmost branch in the tree 
symbolizes the first descent made by DPLL. Above node $G_+$, 
instances are almost surely satisfiable while below
$G_-$, instances have no solutions. The size of the critical
window, that is, the number of variables to assign to reach $G_-$ from
$G_+$, is $W \ll N$. $G_0$ denotes the highest node in the tree carrying a
satisfiable 2+p-SAT instance. A grey triangle accounts for 
(exponentially) large refutation subtree that DPLL has to go through
before backtracking above $G_-$. By definition, the highest
node reached back by DPLL is $G_0$. Further backtracking, below $G_0$, might
be necessary but a solution will be eventually found along the rightmost branch
issued from $G_0$.}
\label{treeinter}
\end{figure}
\end{center}

\begin{center}
\begin{figure}
\includegraphics[width=240pt,height=300pt,angle=-90]{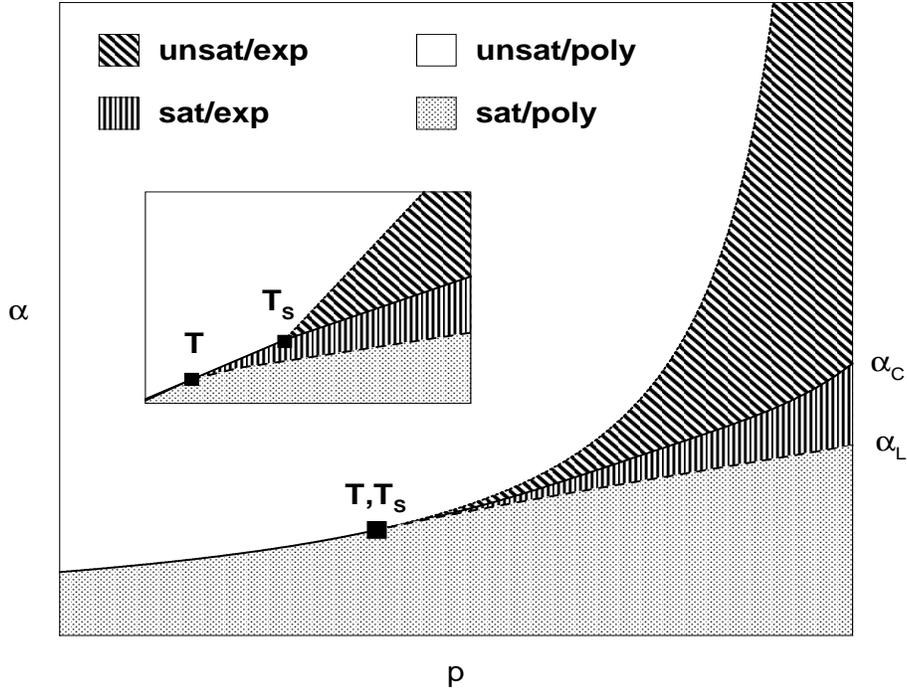}
\vskip .8cm
\caption{Four different regions coexist in the phase diagram of
the 2+p-SAT model, according to whether complexity is polynomial or
exponential, and formulae are sat or unsat. Borderlines are
(from top to down): $\alpha = 1/(1-p)$ (dotted line), $\alpha _C (p)$
(full line), and
the branch trajectory  (dashed line), starting in $(1,\alpha_L)$ and ending
at point T tangentially to the threshold line. 
The tricritical
point T$_S$, with coordinates $p_S\simeq 0.41, \alpha_S= 1/(1-p_S)$, separates
second from first order critical point on the threshold line, and lies
very close to T.
%On the left of the line $\alpha =1/(1-p)$, contradictions 
%immediately arise preventing the search tree from growing.
Inset: schematic blow-up of the T, T$_S$ region (same symbols
as in the main picture).}
\label{pattern}
\end{figure}
\end{center}

In this paper we have analyzed the action of a search algorithm, the DPLL
procedure, on random 3-SAT instances to derive the typical
 complexity as a function of the size (number of variables) $N$ of the
instance and the number $\alpha$ of clauses per variables. 
The easy, polynomial in $N$, as well as the hard,  
exponential in $N$, regimes have been investigated.
We have measured, through numerical simulations, the size and the structure 
of the search tree by computing the number of nodes, the distribution of 
branch lengths, and the highest backtracking point. From a theoretical 
point of view, we have analyzed the dynamical evolution of a randomly
drawn 3-SAT instance under the action of DPLL. The random 3-SAT 
statistical ensemble, described by a single parameter $\alpha$, is not 
stable under the action of DPLL. Another variable $p$, the fraction 
of length three clauses, has to be considered to account for the later 
evolution of the instance\footnote{This situation is reminiscent of what 
happens in real--space renormalization, e.g. decimation. New couplings,
absent in the initial Hamiltonian are generated that must be taken into
account. The renormalization flow takes place in the smallest coupling
space, stable under the decimation procedure, that includes the 
original Hamiltonian as a point (Leo Kadanoff, {\em
private communication}).}. Parameters
$p$ and $\alpha$ are the coordinates of the phase diagram of 
Figure~\ref{diag}. The dynamical evolution of the instance is itself
of stochastic nature, due to the random choices made by the splitting 
rule. We can follow the ensemble evolution in 'time', that is, the number
of  variables assigned by DPLL, and represent this evolution by a 
trajectory in the phase diagram of Figure~\ref{diag}. 
For 3-SAT instances, located on the $p=1$ axis, we show that there 
are three different behaviors, depending on the starting ratio $\alpha$.
In the low sat phase, $\alpha<\alpha_L \simeq 3.003$, trajectories are
always confined in the sat region of the phase diagram. 
As a consequence, the search tree reduces essentially to a simple branch,
and the complexity scales linearly with $N$. On the opposite, 
in the unsat phase, the algorithm has to build a complete search 
tree, with all branches ending with a contradiction, to prove 
unsatisfiability. We have imagined a tree growth process that reflects 
faithfully the DPLL rules for assigning a new literal on a branch, but 
in which all branches evolve in parallel, not in the real backtracking,
sequential way. We have derived a partial differential equation describing 
the stochastic growth of the search tree. The tree trajectory plotted on 
phase diagram of Figure~\ref{diag} represents the evolution of 
the instance parameters $p,\alpha$ for typical, statistically 
dominant branches in the tree.
When the trajectory hits the halt line, contradictions prevent the
tree from further growing. 
Computational complexity is, to exponential order, equal to
the  number of typical branches. Last, in the upper sat phase $\alpha_L
<\alpha<\alpha_c$, the trajectory intersects the critical line $\alpha _C(p)$
in some point G shown in Figure~\ref{diag}, and enters the unsat phase of
2+p-SAT instances. Below G, a complete refutation subtree has to be built. 
The full search tree turns out to be a mixture of a single branch and some
(not exponentially numerous in $N$) complete subtrees 
(Figure~\ref{treeinter}). The exponential 
contribution to the complexity is simply the size of the  subtree that 
can be computed analyzing the growth process starting from G. 

Statistical physics tools can be useful to study the solving complexity 
of branch and bound algorithms\cite{NPC,Hay} applied to hard 
combinatorial optimization or decision problems.  The phase diagram 
of Figure~\ref{diag} affords an accurate understanding of the 
probabilistic complexity of DPLL variants on
random instances. This view may reveal the nature of the complexity of
search algorithms for SAT and related NP-complete problems. In the sat
phase, branch trajectories are related to polynomial time computations
while in the unsat region, tree trajectories lead to exponential
calculations. Depending on the starting point (ratio $\alpha$ of the
3-SAT instance), one or a mixture of these behaviors is observed.
A recent study of the random vertex cover problem \cite{weigt}
has shown that our approach can be successfully applied to other
decision problems.

Figure~4 furthermore gives some insights to improve the search
algorithm. In the unsat region, trajectories must be as horizontal as
possible (to minimize their length) but resolution is necessarily
exponential\cite{Chv}. In the sat domain, heuristics making
trajectories steeper could avoid the critical line $\alpha _C (p)$ and
solve 3-SAT polynomially up to threshold.

Fluctuations of complexity are another important issue that would
deserve further studies. The numerical experiments reported
in Figure~\ref{histoseuil} show that the
annealed complexity, that is, the average solving time required by
DPLL, agrees well with the typical complexity in the
unsat phase but discrepancies appear in the upper sat phase.
It comes as no surprise that our analytical framework, designed 
to calculate the annealed complexity, provides accurate results 
in the unsat regime. We were also able to get rid of the fluctuations,
and to calculate the typical complexity in the 
upper sat phase of 3-SAT from the annealed complexity of critical
2+p-SAT (Table~\ref{tabmixed}). This suggests that fluctuations 
may originate from atypical points G in the mixed structure of the
search tree unveiled in Section~\ref{secmixed}. 
Such atypical points G, coming from the $1/\sqrt N$ finite
size fluctuations of the branch 
trajectory\footnote{Fluctuations also come from the
finite width $W \sim N^{1-1/\nu}$ of the critical 2+p-SAT
line. Recently derived lower bounds on the critical exponent $\nu$
($\ge 2$ \cite{wilson}) reveal that finite size effects could be
larger than $O(1/\sqrt N)$; for 2-SAT indeed, $\nu=3$ and relative
fluctuations scale as $W/N\sim N^{-2/3}$}, lead to exponentially large 
fluctuations of the complexity (Figure~\ref{histoseuil}B).

It would be rewarding to achieve a better theoretical understanding of
such fluctuations, and especially of fluctuations of solving times
from run to run of DPLL procedure on a single instance\cite{Kir}. 
Practitioners of hard problem solving have reached empirical evidence 
that exploiting
in a cunning way the tails of the complexity distribution may allow
a drastic improvement of performances\cite{seltail}.
Suppose you are given one hour CPU time to solve one
instance which, in 99\% of cases, require 10 hours of calculation, and
with probability 1\%, ten seconds only. Then, you could proceed by running
the algorithm for eleven seconds, stop it if the instance has not been
solved yet, start again hundreds of time if necessary till the completion
of the task. Investigating whether such a procedure could be used
to treat successfully huge 3-SAT instances would be very interesting.

\vskip .3cm \noindent
{\bf Acknowledgements:} We are grateful to J. Franco for his
precious encouragements and helpful discussions. 
% We also thank
%A. Hartmann and M. Weigt for their interest in 
%our work\cite{weigt}.
We thank S. Coppersmith and J. Marko for their 
hospitaliy during the final redaction of this paper. R.M. is partly
funded by the ACI Jeunes Chercheurs ``Optimisation combinatoire et
verres de spins quantiques'' from the French Ministry of Research.

\appendix
\section{Largest eigenvalues and eigenvectors of the effective branching
matrix.}
\label{diagosing}

In this appendix, the largest eigenvalue of the effective branching matrix
(\ref{redmat}) is computed. 
We start by multiplying both sides of the eigenvalue equation, obtained by
applying the matrix $\tilde K _{q_2}$ onto the eigenvector
$\tilde{v}_q(C_1')$, by $x^{C_1}$. Then, we sum over $C_1$ and obtain
the following equation for the eigenvectors generating functions
$\tilde V_q(x)$,
\begin{equation}
\label{eiggf3}
\tilde V_q(x)=\tilde V_q(0) \frac{ L(x) \; N(x)}
{{\Lambda}_q- L(x)} \quad .
\end{equation}
where 
\begin{equation}
N(x) = e^{- y_2} \;(x+x^2 )- 1 \qquad , \qquad
L (x) = \frac 1x \; \exp\left( \frac {m_2} {2} \, e^{- y_2} \, x\right) 
\qquad ,
\end{equation}
and
\begin{equation}
\Lambda_{q} = \tilde{\lambda}_{q , q_2} \exp \left\{ m_2\, \left(
1 - \frac{e^{-y_2}}{2} \right) \right\} \qquad ,
\label{eiggf}
\end{equation}
with $y_2 = - i q_2$. From  Section~\ref{sectree}, $q_2$ is 
purely imaginary at 
saddle-point, and it is therefore convenient to manipulate the real-valued 
number $y_2$. 

\subsection{Zeroes and poles of $\tilde V_q$.} \label{pole}

\begin{center}
\begin{figure}
\includegraphics[width=220pt,height=300pt,angle=-90]{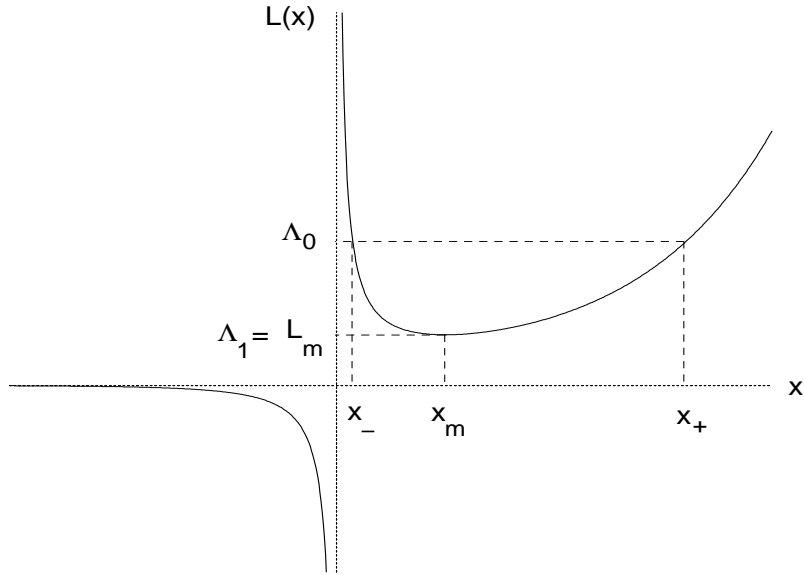}
\vskip .3cm
\caption{Sketch of the function $L(x)$ 
appearing in the denominator of the eigenvector
generating function $\tilde V_q(x)$. 
$L(x)$ is positive (resp. negative) for
positive (resp. negative) arguments $x$. The local positive minimum is
located at $x_m= 1/\gamma _2, L_m = e\, \gamma _2$. The height of the 
minimum, $L_m$, is equal to the edge $\Lambda_1$ of the (excited states)
continuous spectrum. For $\Lambda > L_m$, the equation $L(x)=\Lambda$ has 
two roots $x_-, x_+$ such that $x_- < x_m < x_+$. When $x_-$ coincides
with the positive zero $x^*$ of the numerator $N(x)$,
the maximal eigenvalue $\Lambda _0$ is obtained (Appendix~\ref{diagosing}).}
\label{graphdeL}
\end{figure}
\end{center}

$N(x)$ has two zeroes $x^\dagger < 0 < x ^*$ that are
functions of $y_2$ solely and given by
\begin{eqnarray}
x^ \dagger &=& \frac 12 \left( -1 - \sqrt{ 1 + 4 e^{y_2}} \right) \nonumber \\
x^*  &=& \frac 12 \left( -1 + \sqrt{ 1 + 4 e^{y_2}} \right) \qquad .
\label{xstar}
\end{eqnarray}
$N(x)$ is negative when its argument $x$ lies between the zeroes, positive 
otherwise.
The function $L(x)$ is plotted Figure~\ref{graphdeL}. The positive local minimum
of $L(x)$ is located at $x_m =2\; e^{y_2} /m_2, L_m = m_2 
\, e^{1- y_2} /2 $.   
The number of poles of $\tilde V_q(x)$ can be inferred from 
Figure~\ref{graphdeL}. 
\begin{itemize}
\item If $ {\Lambda}_q < 0$, there is a single negative pole.
\item If $ 0 \le {\Lambda}_q < L_m$, there is no pole.
\item If $ {\Lambda}_q > L_m$, there are two positive
poles $x_-, x_+$ with $x_- < x_m < x_+$ that coalesce when 
${\Lambda}_q = L_m$.
\end{itemize}

\subsection{Largest eigenvalue and eigenvector.} \label{largest}

Consider the largest eigenvector $\tilde
v _0 (C_1)$ and the associated eigenvalue $\Lambda _0 > 0$. The ratios 
\begin{equation}
\mu (C_1)  = \frac{ v _0 (C_1)} {\sum _{C'_1 =0} ^\infty   v _0 (C'_1)}
\end{equation}
define the probability that the number of unit-clauses be equal to $C_1$ at 
a certain stage of the search. Consequently, as long as no contradiction
occurs, we expect all the ratios to be positive and decrease quickly with 
$C_1$. The generating function $\tilde V_q(x)$ must have a finite radius
of convergence $R$, and be positive in the range $0\le x< R$. 
Note that the radius of convergence coincides with a pole of $\tilde V_q$.

The asymptotic behavior of $\tilde v _0 (C_1)$ is simply given by the radius
of convergence,
\begin{equation}
\tilde v _0 (C_1) \sim \left( \frac 1R \right) ^{C_1} \qquad ,
\end{equation}
up to non-exponential terms. New branches
are all the more frequent that many splittings are necessary and unit-clauses
are not numerous, {\em i.e.} $\tilde
v_0 (C_1)$ decreases as fast as possible.
From the discussion of Section~\ref{pole}, the radius of convergence is
maximal if $R=x_+$. To avoid the presence of another pole to $\tilde
V _q(x)$ in $x_-$, the zero of the numerator function $N(x)$
must fulfills $x^*=x_-$. 
We check that $\tilde V_q (x)$ is positive for $0\le x < R$. The 
corresponding
eigenvalue is $\Lambda _0 = L(x_+)$, see eqn.~(\ref{eiggf2}).
In next Section, we explain that this theoretical value of $\Lambda _0$ has 
to be modified in a small region of the $(y_2,m_2$) plane. 

The eigenvector $\tilde v_0$ undergoes a delocalization transition 
on the critical line $x_+ (y_2,m_2)=1$ or, equivalently, $L(x^*)=
L(1)$. At fixed $y_2$, the eigenvector is 
localized provided that parameter $m_2$ is smaller than a critical value
\begin{equation}
m_2 ^{loc. 0} ( y_2) = 2 \; e^{y_2} \;\frac{ \ln x^* (y_2)}{  x^* (y_2) -1}
\qquad . 
\label{m20deloc}
\end{equation}
The corresponding curve is shown in Figure~\ref{m2critcourbe}. 

\subsection{Excited state and spectral gap.}

As $\tilde K_{q_2}$ is not a symmetric matrix, complex 
eigenvalues may be found in
the spectrum. We have performed numerical investigations by diagonalizing
upper left corners of $\tilde K_{q_2}$ of larger and larger sizes. From a 
technical point of view, we define the $U\times U$ matrix $\tilde
K ^{(U)} _{q_2}(C_1 ,
C'_1) = \tilde K_{q_2} (C_1,C'_1)$ for $0\le C_1 , C'_1 \le U$. Numerics
show that complex eigenvalues are of small modulus with respect to real-valued
eigenvalues. 

If $y_2 < \ln 2$,
the largest eigenvalue $\Lambda ^{(U)} _0$ of $\tilde
K ^{(U)} _{q_2}$ converges very quickly to
the theoretical value $L(x^+)$ as $U$ increases (with more than 
five correct digits when $U \ge 30$). 
At small values of $m_2$, the second largest
eigenvalue (in absolute value) is negative. Let us call it $\Lambda _
\dagger$. The associated eigenvector $\tilde v_\dagger (C_1)$ is localized; 
all components of $\tilde 
v_\dagger$ have the same sign except $\tilde v_\dagger (0)$. The 
value of $\Lambda _\dagger$ may be computed along the lines of 
Section~\ref{largest},
\begin{equation}
\Lambda _\dagger = L ( x^\dagger ) \qquad .
\end{equation}
Indeed, $x^\dagger < 0 $ implies from Figure~\ref{graphdeL} 
that $L(x^\dagger) < 0$.  
As $m_2$ increases (at fixed $y_2$), $\Lambda _\dagger$ becomes smaller (in
modulus) than the second largest positive eigenvalue $\Lambda_1$. $\Lambda_1$
is followed by a set of positive eigenvalues $\Lambda _2 > \Lambda _3 > 
\ldots$. Successive $\Lambda _q$ ($q\ge 1$) gets closer and closer as 
$U$ increases, to form a continuous spectrum in the $U\to \infty$ limit. 

\begin{center}
\begin{figure}
\includegraphics[width=250pt,height=300pt,angle=-90]{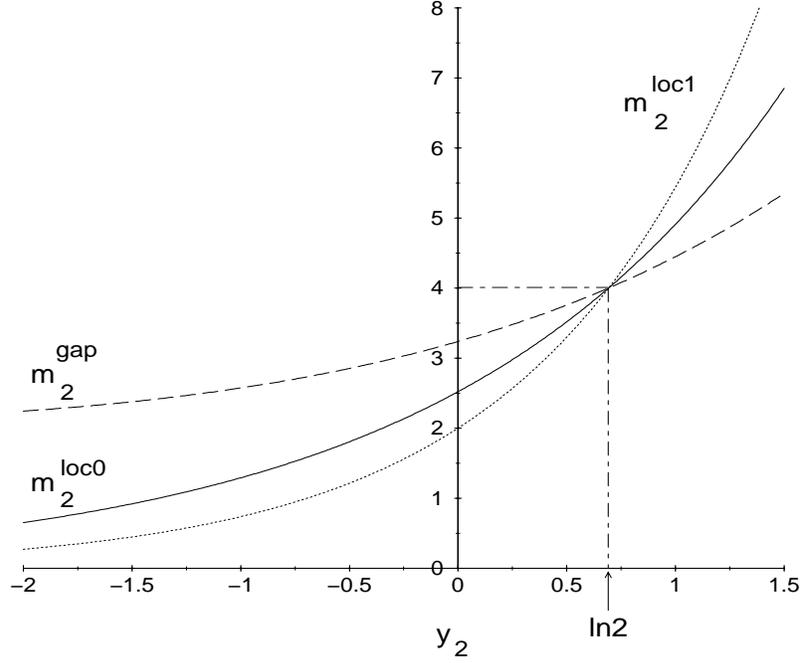}
\vskip .3cm
\caption{Critical curves of $m_2=2 c_2/(1-t)$ as a function of the
parameter $y_2$. From bottom to top (left side): delocalization
threshold $m_2 ^{loc.1}$ for the second largest eigenvector (dotted
line), delocalization threshold $m_2 ^{loc.0}$ for the largest
eigenvector (full line), and zero gap curve $m_2 ^{gap}$ (dashed line).
All curve meet in $y_2 = \ln 2, m_2=4$. For $y_2 < \ln 2$
and small $m_2$, the largest eigenvector is separated from a 
continuum of excited states by a finite gap,
and undergoes a delocalization condition when $m_2$ reaches $m_2 ^{loc.0}$.
For $y_2 > \ln 2$, the largest eigenvector merges with the
continuum spectrum when $m_2 \ge m_2 ^{gap}$,
and gets delocalized on the critical line $m_2 ^{loc1}$.}
\label{m2critcourbe}
\end{figure}
\end{center}

The eigenvectors $\tilde v _q$ ($q=1,2,3,\ldots)$ have real-valued 
components of periodically changing signs. The corresponding generating 
functions have therefore complex-valued radii of convergences, and $V_q (x)$
does not diverge for real arguments $x$. Consequently, the  
edge of the continuous spectrum $\Lambda_1$ is given by
\begin{equation}
\Lambda _1= L _{m} .
\label{exprlam1}
\end{equation}
The above theoretical prediction is in excellent agreement with 
the large $U$ extrapolation of the numerical values of $\Lambda _1$.
Repeating the discussion of Section~\ref{largest}, the first excited state
$\tilde v_1$ becomes delocalized when $x_m=1$, that is, when 
$m_2$ exceeds the value
\begin{equation}
m_2 ^{loc. 1} ( y_2) = 2 \; e^{y_2} 
\qquad . 
\label{m21deloc}
\end{equation}
The corresponding curve is shown in Figure~\ref{m2critcourbe}. 

The gap between $\Lambda _0$ and $\Lambda _1$ will be strictly positive as
long as $x^*<x_{m}$. This defines another upper value for $m_2$,
\begin{equation}
m_2 ^{gap} ( y_2) = 2 \; e^{y_2} \frac 1{x^*(y_2)} 
\qquad ,
\label{m2gap}
\end{equation}
beyond which the largest eigenvalue coincides with the top $\Lambda _1$
of the continuous spectrum. As can be seen from Figure~\ref{m2critcourbe}, 
$\Lambda _0$ coincides with $L(x^+)$
in the region $m_2 < m_2 ^{loc0}$ as long as the largest eigenvalue
 is separated from the 
continuous spectrum by a finite gap. Therefore, in the region
$(y_2 > \ln 2, m_2 ^{gap} < m_2 < m_2 ^{loc0})$, the largest eigenvalue 
merges with $\Lambda _1$, and is given by eqn. (\ref{exprlam1}).

When $m_2$ crosses the critical line $m_2 ^{loc0}$, the largest
eigenvector gets delocalized and the average number 
of unit-clauses flows to infinity. As a result of the avalanche of unitary 
clauses, contradictions necessarily occur and the growth of the tree stops.
Notice that, as far as the total number of branches is concerned, we shall
be mostly concerned by the $y_2=0$ axis. The critical values of interest are
in this case: $m_2 ^{loc1}=2$, $m_2 ^{loc0}=(3+\sqrt 5) \ln[(1+\sqrt 5)/2]
\simeq 2.5197$ and $m_2 ^{gap}=1+\sqrt 5 \simeq 3.2361$.

%%%%%%%%%%%%%%%%%%%%%%%%

\section{Quadratic approximation for the growth partial differential equation.}
\label{quadapp}

The partial differential equation for the growth of the search tree may be
written in terms of the Legendre transform $\varphi (y_2 , y_3 , t)$ of the 
logarithm of the number of branches as
\begin{equation}
\frac{\partial \varphi}{\partial t} (y_2 , y_3 , t) = g_1 (y_2) - \frac 2{1-t}
g_2 (y_2 ) \frac{\partial \varphi}{\partial y_2}- \frac 3{1-t}
g_3 (y_2 , y_3) \frac{\partial \varphi}{\partial y_3}
\quad ,
\label{pde1}
\end{equation}
with
\begin{eqnarray}
g_1 (y_2) &=& -y_2 + \ln \left[ \frac 12 \left( 1 + \sqrt{ 1+ 4 e^{2 y_2}}
\right) \right] \nonumber \\
g_2 (y_2) &=& 1 - \frac 12 \left[ 1 + \frac 12 \left( - e^{-y_2} +
\sqrt{ 4 + e^{-2 y_2}} \right) \right] \nonumber \\
g_3 (y_2 , y_3) &=& 1 - \frac 12 e^{-y_3} \left( 1+ e^{y_2} \right)
\qquad . \label{fonctionsg}
\end{eqnarray}

\subsection{Linear approximation.}

At the first order in $y_2, y_3$, we replace the functions $g$ 
appearing in (\ref{pde1}) with their linearized counterparts, 
\begin{eqnarray}
g_1 ^{(1)} (y_2) &=& \ln \left( \frac{ 1 +\sqrt 5}{2} \right) - 
\left( \frac { 5 +\sqrt 5} {10} \right) \; y_2 \nonumber \\
g_2 ^{(1)}(y_2) &=&  \frac{ 3 -\sqrt 5} 4  +
\left( \frac{ 5+ 3 \sqrt 5} 4 \right) y_2 \nonumber \\
g_3 ^{(1)} (y_2 , y_3) &=& y_3 - \frac 12 y_2
\qquad , \label{fonctionsglin}
\end{eqnarray}
and solve the corresponding partial differential equation. The 
solution, called $\varphi ^{(1)} (y_2 , y_3 ,t)$ is given in equations 
(\ref{solulin89}) and (\ref{c3c}).

\subsection{Quadratic approximation.}

At the second order in $y_2, y_3$, we consider the quadratic corrections
to the $g$ functions,
\begin{eqnarray}
g_1 ^{(2)} (y_2) &=& \frac {\sqrt 5} {50} ( y_2 )^2 \nonumber \\
g_2 ^{(2)}(y_2) &=& - \left( \frac{ 25 +11 \sqrt 5}{ 200} \right) (y_2)^2
\nonumber \\
g_3 ^{(2)} (y_2 , y_3) &=& -\frac 12 (y_3)^2 + \frac 12 y_3 y_2 -\frac 14
(y_2)^2 \qquad , \label{fonctionsgquad}
\end{eqnarray}
and look for a solution of (\ref{pde1}) of the form $\varphi = \varphi ^{(1)}
+ \varphi ^{(2)}$. Neglecting higher order terms in $g ^{(2)} \varphi ^{(2)}$,
we end with
\begin{equation}
\frac{\partial \varphi ^{(2)} }{\partial t} (y_2 , y_3 , t) = G ^{(2)} 
(y _2, y_3)  - \frac 2{1-t}
g_2 ^{(1)} (y_2 ) \frac{\partial \varphi ^{(2)}}{\partial y_2}- \frac 3{1-t}
g_3 ^{(1)}(y_2 , y_3) \frac{\partial \varphi^{(2)} }{\partial y_3}
\quad ,
\label{pde2}
\end{equation}
where
\begin{equation}
G ^{(2)} (y _2, y_3) = g_1 ^{(2)} (y_2) - \frac 3{1-t}
g_2 ^{(2)} (y_2 ) \frac{\partial \varphi ^{(1)}}{\partial y_2}- \frac 2{1-t}
g_3 ^{(2)}(y_2 , y_3) \frac{\partial \varphi^{(1)} }{\partial y_3}
\quad .
\label{foncG2}
\end{equation}
A particular solution of (\ref{pde2}) may be found under the form
\begin{equation}
\varphi ^{(2)} _{part.} (z_2 , z_3 , t) = a _0 (t) + a_{21} (t) y_2 +
a_{22} (t) (y_2)^2 + a_{33} (t) (y_3)^2 \quad ,
\end{equation}
where the $a$'s 
are linear combinations of $1-t$, $(1-t)^3$ and $(1-t)^\mu$ with
$\mu=(5+3\sqrt 5)/10$. 

The general solution of the homogeneous version of (\ref{pde2}) reads
\begin{equation}
\varphi ^{(2)} _{hom.} (z_2 , z_3 , t) = \Phi \bigg[ (1-t)^3 \left(y_3-
\frac{75+9\sqrt 5}{116} y_2 - \frac {15-4\sqrt 5}{58} \right) ;
(1-t) ^\mu \left( y_2 +\frac {7\sqrt 5 -15}{2}\right ) \bigg] \ ,
\end{equation}
where $\Phi$ is a differentiable function of two arguments $u,v$.
Assuming that $\Phi [u;v]$ is a quadratic form in $u$ and $v$, we fix its
six coefficients through the initial condition (at time $t=0$),
\begin{equation}
\varphi ^{(2)} (y_2 , y_3 , 0) = \varphi ^{(2)} _{part.}  (y_2 , y_3 , 0)
+ \varphi ^{(2)} _{hom.}  (y_2 , y_3 , 0) = 0 \ . 
\end{equation}
The resulting expression for $\varphi ^{(2)}$ is too long to be given here but
can be obtained safely by using an algebraic computation software, e.g.
Mathematica.

\newpage
\begin{center}
{\large TABLES}
\end{center}

\begin{table}
$$
\begin{array}{|c |c |c| c|}
\hline 
{\hbox{\rm Ratio}}\;  \alpha \;{\hbox {clause/var}} & 
{\hbox{\rm Nb. of var.}} & 
{\hbox{Resolution time}} \\ 
\hline \hline
20  & 900 & 6\;{\hbox{ seconds}}  \\
10  & 700 & 1\;{\hbox{ hour}}  \\
7   & 400 & 20\;{\hbox{ minutes}} \\
4.3 & 350 & 2\;{\hbox{ days}}  \\
3.5 & 500 & 20\;{\hbox{ minutes}}  \\
\hline\hline
\end{array}
$$     
\caption{Typical computational time required to  
solve some hard random 3-SAT instances for different
ratios $\alpha$ and sizes $N$. Slightly below
threshold, some rare instances may require much longer computational
time (Section~\ref{foc}).}
\label{tabtemps}
\end{table}

\begin{table}
$$
\begin{array}{|c |c| c|c| c|}
\hline 
\multicolumn{1}{|c}{\hbox{\rm Ratio } \ \alpha} & 
\multicolumn{2}{|c}{\hbox{\rm Experiments}} & 
\multicolumn{2}{|c|}{\hbox{\rm Theory}} \\
{\hbox{\rm of clause/var.}} & {\hbox{\rm nodes}}&{\hbox{\rm histogram }}& 
{\hbox{\rm lin.}} & {\hbox{\rm quad.}} \\ 
\hline \hline
20& 0.0153 \pm 0.0002 & 0.0151 \pm 0.0001  & 0.0152  & 0.0152 \\
15& 0.0207 \pm 0.0002 & 0.0206 \pm 0.0001  & 0.0206  & 0.0206 \\
10& 0.0320 \pm 0.0005 & 0.0317 \pm 0.0002  & 0.0319  & 0.0319 \\
7 & 0.0481 \pm 0.0005 & 0.0477 \pm 0.0005  & 0.0477  & 0.0476 \\
4.3& 0.089 \pm 0.001  & 0.0895 \pm 0.001   & 0.0875  & 0.0852 \\
\hline \hline
\end{array}
$$

\caption{Logarithm of the complexity $\omega$ from experiments and theory in 
the unsat phase.
Values of $\omega$ from measures of search tree sizes (nodes),
histograms of branch lengths (histogram) and
theory within linear (lin.) and quadratic (quad.)
approximations. Note the excellent agreement between theory
and experiments for large $\alpha $.}

\label{tabunsat}
\end{table}

\begin{table}
$$
\begin{array}{|c | c|c|c|}
\hline 
\multicolumn{1}{|c}{\hbox{\rm Ratio } \ \alpha _0} & 
\multicolumn{1}{|c}{\hbox{Slope}} & 
\multicolumn{2}{|c|}{\hbox{Curvature} \ \beta}\\
{\hbox{\rm of clause/var.}} & \gamma &
{\hbox{\rm nodes}}&{\hbox{\rm histogram }} \\ 
\hline \hline
15 & -1.47 & 69.6 & 75.6 \\
10 & -1.32 & 56.5 & 47.8 \\
7  & -1.06 & 39.4 & 29.6 \\ 
4.3 & -0.58 & 20.2& 13.6 \\
\hline\hline
\end{array}
$$
\caption{Details on the fits of search tree sizes from equation 
(\ref{nodeq}).
For different ratios $\alpha$, the slope $\gamma$ of the fit of 
$\omega _N - \log _2 N/N$
vs. $1/N$ is shown as well as the corresponding curvature
$\beta$ deduced from (\ref{nodeq}) (column ``nodes''). 
The curvatures measured
directly at the top of the branch lengths histograms are listed in the
``histogram'' column.
}
\label{tabfit}
\end{table}

\begin{table}
$$
\begin{array}{|c| c|c|c| c|c| c|}
\hline 
\multicolumn{2}{|c}{\hbox{\rm Parameters}  } & 
\multicolumn{3}{|c}{\hbox{\rm Experiments}} & 
\multicolumn{2}{|c|}{\hbox{\rm Theory}} \\
p & \alpha  & \omega ^{ann} & \omega ^{typ}_ {nod.}&
\omega ^{typ}_ {his.}& {\hbox{\rm lin.}} & {\hbox{\rm quad.}} \\ 
\hline \hline
 1   & 3.5    & 0.043 \pm 0.002 & 0.035 \pm 0.003 & & 0.0355 & 0.0329  \\
0.78 & 3.02   & 0.044 \pm 0.002 & 0.041 \pm 0.002 & 0.041 \pm 0.002 & 
0.0440 & 0.0407 \\
\hline \hline
\end{array}
$$

\caption{Logarithm of the complexity $\omega$ from experiments 
and theory in the upper sat phase.
Experiments determine $\omega$ from measures of the annealed complexity
(ann.), of the typical search tree sizes (nod.) and of 
histograms of branch lengths (his.). Data are presented for 3-SAT instances
with ratio $\alpha =3.5$, and  2+p-SAT instances
with parameters $p=0.78, \alpha = 3.02$.
Theoretical predictions within linear (lin.) and quadratic (quad.)
approximations are reported for the 2+p-SAT model, and for 3-SAT
using eqn.(\ref{erreur}).}

\label{tabmixed}
\end{table}

\end{document}